\newif\ifarxiv\arxivtrue

 \arxivtrue     

\ifarxiv
\documentclass[12pt,a4paper]{article}
 \pdfoutput=1
\usepackage{titling} 
\usepackage[labelfont=bf]{caption}

\newcommand\afterTocSpace{\bigskip\medskip}
\newcommand\afterTocRuleSpace{\bigskip\bigskip}

\makeatletter

\divide\@tempdima\baselineskip
\@tempcnta=\@tempdima
\makeatother

\else
\documentclass{pinchcrn}   
\fi

\usepackage{mathtools,amsmath,amssymb}    
\usepackage{hyperref}
\usepackage{subcaption}
\usepackage{simplewick}
\usepackage{dsfont}

\usepackage[text
size=footnotesize]{todonotes}
\makeatletter
\DeclareRobustCommand*{\bfseries}{%
  \not@math@alphabet\bfseries\mathbf
  \fontseries\bfdefault\selectfont
  \boldmath
}

\makeatother

\usepackage{graphicx}

\usepackage{feynmp}

\usepackage{ifpdf}

\ifpdf
\else
\fi

\ifarxiv

\title{Introduction to Integrability and One-point Functions in \texorpdfstring{${\mathcal{N}=4}$}{N=4} SYM  and
its Defect Cousin}
\author{M.\ de Leeuw, A.C.\ Ipsen, C.\ Kristjansen and M.\ Wilhelm}

\else

\title{Guidelines for preparing a manuscript using \LaTeX\ and
the Oxford University Press {\tttitle pinchcr} class file}

\author{M.\ de Leeuw, A.C.\ Ipsen, C.\ Kristjansen\footnote{Lecturer at the school.} and M.\ Wilhelm}
\affiliation{Niels Bohr Institute, Copenhagen University \\
Blegdamsvej 17, DK-2100 Copenhagen \O}
\authors{2}

\fi

\DeclareMathOperator{\idm}{\mathds{1}}

\newcommand{\de}{\operatorname{d}\!}
\newcommand{\e}{\operatorname{e}}
\newcommand{\eqndot}{\, . }
\newcommand{\eqncom}{\, , }

\newcommand{\MPS}{\mathrm{MPS}}
\newcommand{\Neel}{\mathrm{N\acute{e}el}}
\newcommand{\tr}{\mathrm{tr}}
\newcommand{\Obare}{\mathcal{O}^{\text{bare}}}
\newcommand{\Oren}{\mathcal{O}^{\text{ren}}}
\newcommand{\Obarebar}{\bar{\mathcal O}^{\text{bare}}}
\newcommand{\Orenbar}{\bar{\mathcal O}^{\text{ren}}}
\newcommand{\Mbare}{\tilde{M}}
\newcommand{\logMbare}{\mathcal{M}}
\newcommand{\YM}{{\mathrm{\scriptscriptstyle YM}}}
\newcommand{\renZ}{\mathcal{Z}}

\DeclareMathOperator{\cder}{D}
\newcommand{\scal}{\phi}
\newcommand{\scalc}{\scal^{\text{cl}}}
\newcommand{\scalq}{\tilde{\scal}}
\newcommand{\ferm}{\psi}
\newcommand{\aferm}{\bar{\ferm}}

\newcommand{\Xc}{X^{\text{cl}}}
\newcommand{\Yc}{Y^{\text{cl}}}

\newcommand{\Yq}{\tilde{Y}}
\newcommand{\Xq}{\tilde{X}}

\newcommand{\comm}[2]{[#1,#2]}
\newcommand{\acomm}[2]{\{#1,#2\}}
\newcommand{\cO}{\mathcal{O}}

\newcommand{\gammaE}{\gamma_{\text{E}}}
\newcommand{\peps}{\varepsilon}

\renewcommand{\digamma}{\Psi}

\newcommand{\ihalf}{\frac{i}{2}}

\newcommand{\permop}{\mathbb{P}}
\newcommand{\traceop}{\mathbb{K}}

\DeclareMathOperator{\phaneq}{\phantom{{}=}}

\usepackage[most]{tcolorbox}
\definecolor{block-gray}{gray}{0.85}
\newtcolorbox{exc}{colback=block-gray,
boxrule=0pt,boxsep=0pt,breakable}

\begin{document}
\begin{fmffile}{diagrams}


\ifarxiv

\begingroup\parindent0pt
\begin{flushright}\footnotesize
\end{flushright}
\vspace*{4em}
\centering
\begingroup\LARGE
\bf
Introduction to Integrability and One-point Functions in \texorpdfstring{${\mathcal{N}=4}$}{N=4} SYM  and
its Defect Cousin\par\endgroup
\vspace{2.5em}
\begingroup\large{\bf Marius de Leeuw, Asger C.\ Ipsen, Charlotte Kristjansen,\\ and Matthias Wilhelm}
\par\endgroup
\vspace{1em}
\begingroup\itshape
Niels Bohr Institute, Copenhagen University,\\
Blegdamsvej 17, 2100 Copenhagen \O{}, Denmark\\

\par\endgroup
\vspace{1em}
\begingroup\ttfamily
deleeuwm@nbi.ku.dk, asgercro@nbi.ku.dk, kristjan@nbi.ku.dk, matthias.wilhelm@nbi.ku.dk \\
\par\endgroup
\vspace{2.5em}
\endgroup

\begin{abstract}
\noindent
These lectures give a basic introduction to $\mathcal{N}=4$ SYM theory and the integrability of its planar spectral problem  as seen from the perspective of a recent development, namely the application of  integrability techniques in the study of one-point functions in a defect version of the theory.
\end{abstract}

\bigskip\bigskip\par\noindent

\thispagestyle{empty}

\newpage
\hrule
\tableofcontents
\afterTocSpace
\hrule
\afterTocRuleSpace

\else

\maketitle

\fi

\ifarxiv
\else
\chapter{Introduction to Integrability and One-point Functions in \texorpdfstring{${\mathcal{N}=4}$}{N=4} SYM  and
its Defect Cousin}
\fi

\section{Introduction}

 $\mathcal{N}=4$ super Yang-Mills (SYM) theory is a distinguished quantum field theory carrying the maximal amount of supersymmetry for a non-gravitational theory in four dimensions and being conformal even at the quantum level. 
 It plays the role
of the CFT in the AdS/CFT correspondence and it exhibits integrability at the planar level.
As a matter of fact, since its formulation almost 40 years ago~\cite{Gliozzi:1976qd,Brink:1976bc} the theory has continuously revealed novel interesting features
and keeps on doing so.   

At the present time, the fundamentals of ${\mathcal N}=4$ SYM theory are treated in several text books such as \cite{Ammon:2015wua,Wess:1992cp},  and there already exists a number of
reviews discussing the integrability of  the theory's planar   spectral problem~\cite{Beisert:2004ry,Minahan:2006sk,Beisert:2010jr,Serban:2010sr}. 
In these lectures, we will be brief about the basics of ${\mathcal N}=4$ SYM theory and 
choose 
a slightly different perspective on its integrability properties, showing
how the tools of integrability inherited from the planar spectral problem can be used to study one-point functions in a certain
defect version of $\mathcal{N}=4$ SYM theory.   
For a discussion of the role of ${\mathcal N}=4$ SYM theory in the AdS/CFT correspondence, we refer to the
lectures by G.\ Semenoff.

We start in section~\ref{ActionandSymmetries} by presenting the action of  ${\mathcal N}=4$ SYM theory and briefly reviewing its symmetries. This allows us to
introduce the key concept of the dilatation operator, the diagonalisation of which constitutes the above-mentioned spectral
problem. Furthermore, having at hand the
explicit expressions for the symmetry generators will facilitate the discussion of symmetry breaking for the defect version
of the theory. 

Next, we move on to discussing in section~\ref{spectrum}  the  two-point functions  of ${\mathcal N}=4$ SYM theory.  By extracting the logarithmically
divergent pieces of the two-point functions, one can read off the dilatation generator of the theory.  For the analysis of (quantum) one-point functions, however,
one needs not only the logarithmically divergent pieces but also the finite parts of the two-point functions.
Hence, we have chosen to present in quite some detail the perturbative calculation of two-point functions in the
scalar sector of the theory, whereby, in addition, we fill  some gaps in the earlier reviews.  

 The dilatation operator of ${\mathcal N}=4$ SYM theory can be identified with the Hamiltonian of an 
integrable spin chain, and at one-loop order specialising to the simplest possible sector of the theory this spin chain  reduces to the Heisenberg spin chain~\cite{Minahan:2002ve}. 
The Hamiltonian of the Heisenberg spin chain can be diagonalised either by coordinate-space or algebraic Bethe ansatz techniques. These techniques were explained in detail in the lectures by J.L.\ Jacobsen.
Here, we will only highlight in section~\ref{spinchain}
some features of the solution which will be of importance for the study of one-point functions, 
namely the parity properties of the eigenstates and their so-called Gaudin norm~\cite{Gaudin:1976sv}.
 In addition, we will discuss, on a general
level, how the spin-chain picture of $\mathcal{N}=4$ SYM theory generalises to higher loop orders. Finally, we will summarise various
observations concerning the non-planar spectral problem.

When defects or boundaries are introduced in a conformal field theory such as $\mathcal{N}=4$ SYM theory, novel features emerge.
Hence, the theory can have non-trivial one-point functions, and two-point functions between operators with unequal
conformal dimension need no longer vanish. There exists a certain defect version of $\mathcal{N}=4$ SYM theory in which some
of the scalar fields acquire a vacuum expectation value characterised by a certain representation label $k$ and where one-point functions are non-trivial already at tree level. This defect conformal field theory (dCFT), moreover, has a holographic dual.
The holographic dual consists of a so-called D5-D3 probe-brane system where a single probe D5 brane with 
geometry $AdS_4\times S^2$ is embedded in the
usual $AdS_5\times S^5$ background of AdS/CFT and carries $k$ units of background gauge field flux on the $S^2$.
For details, we refer to the lectures by G.\ Semenoff.

The remaining part of the lectures will be devoted to the study of this defect version of $\mathcal{N}=4$ SYM theory.
First, in section~\ref{sec: introtoDCFT}
we will analyse its symmetry properties
making explicit the surviving part of the $\mathcal{N}=4$ SYM  symmetry algebra. Subsequently,
we will demonstrate how the tools of integrability can be applied to the calculation of  the one-point functions of the dCFT, first at tree level in section~\ref{tree-level-onepoint} and subsequently in section~\ref{one-loop-one-point} 
at one-loop order. In particular, we will
derive a closed expression for the one-point functions in the simplest so-called $SU(2)$ sector of ${\mathcal{N}=4}$ SYM theory
valid for any operator and for any value of the representation label $k$. 
For the tree-level calculation, the transfer matrix
of the Heisenberg spin chain will be shown to play a crucial role, and for the one-loop calculation we will make use of
our explicit quantum-field-theoretical computations in section~\ref{spectrum}.  We will also briefly mention a proposal for
an  all loop so-called asymptotic one-point function formula. This formula correctly encodes all available one-loop data but
whether or not the formula remains exhaustive  at higher loop orders
constitutes an open question.

 We conclude our lectures by a discussion in section~\ref{Discussion} of this as well as other open questions related to the defect version of 
 ${\mathcal{N}=4}$ SYM theory and in addition briefly list other 
 recent applications of integrability in the context of ${\mathcal{N}=4}$ SYM theory.

Throughout the lecture notes, exercises are provided in order to help the interested student 
 acquiring  some hands-on knowledge of the different concepts.

\section{\texorpdfstring{$\mathcal{N}=4$}{N=4} SYM theory and the spectral problem}


\subsection{Action and symmetries \label{ActionandSymmetries}}

\subsubsection{Action}

The maximally supersymmetric $\mathcal{N}=4$ SYM theory in four dimensions can be constructed from $\mathcal{N}=1$ SYM theory in ten dimensions via dimensional reduction. In this reduction, the ten-dimensional gauge fields splits into the four-dimensional gauge field $A_\mu$, $\mu=0,1,2,3$, and the six real scalars $\phi_i$, $i=1,2,3,4,5,6$. Similarly, the ten-dimensional Majorana-Weyl fermion $\psi$ splits into four four-dimensional Majorana fermions.

The action of $\mathcal{N}=4$ SYM theory reads
\begin{multline}
 \label{eq: SYM-action}
  S_{{\mathcal N}=4}=\frac{2}{g_\YM^2}\int \de^4x\,\tr\biggl[ -\frac{1}{4}F_{\mu\nu}F^{\mu\nu}-\frac{1}{2}\cder_\mu\scal_i\cder^\mu\scal_i\\+\frac{i}{2}\aferm\Gamma^\mu\cder_\mu\ferm +\frac{1}{2}\aferm\Gamma^i\comm{\scal_i}{\ferm}+\frac{1}{4}\comm{\scal_i}{\scal_j}\comm{\scal_i}{\scal_j}\biggr]\eqncom
\end{multline}
where the field strength $F_{\mu\nu}$ and the covariant derivatives $\cder_\mu$ are defined via 
\begin{equation}
 \begin{aligned}
  F_{\mu\nu}&=\partial_\mu A_\nu-\partial_\nu A_\mu-i\comm{A_\mu}{A_\nu}\eqncom\\
  \cder_\mu\scal_i&=\partial_\mu\scal_i-i\comm{A_\mu}{\scal_i}\eqncom \hspace{0.5cm}
  \cder_\mu\psi=\partial_\mu\psi-i\comm{A_\mu}{\psi}\eqndot
 \end{aligned}
\end{equation}
Here, $\Gamma$ denotes the ten-dimensional gamma matrices which govern the coupling of the ten-dimensional fermion $\psi$ to the bosons.
Exact expressions for $\Gamma$ and the reductions of $\Gamma$ and $\psi$ to four dimensions can be found in \cite{Buhl-Mortensen:2016jqo} in our conventions; for the present discussion, they are however not required.

We consider $\mathcal{N}=4$ SYM theory with gauge group $U(N)$. 
All fields transform in the adjoint representation of the gauge group. 
We denote the colour components of, say, the scalars $\phi^i$ as $[\phi^i]_{ab}$, where $a,b=1,\dots, N$ are fundamental indices. 
We can build gauge-invariant local composite operators by taking traces of products of fields that transform covariantly under the gauge group.%
\footnote{For the gauge fields, the covariant combinations are the field strength and covariant derivatives that can act on all fields.}
Moreover, we can take products of such single-trace operators to obtain multi-trace operators.

Mostly, we are restricting ourselves to the 't Hooft limit, where $g_\YM\to 0$, $N\to\infty$ while $\lambda= g_\YM^2 N$ is kept fixed~\cite{tHooft:1973alw}.
In this limit, only planar Feynman diagrams contribute to correlation functions, which is why it is also called the planar limit. 
Moreover, interactions that lead to splitting and joining of traces are suppressed, such that it is sufficient to look at operators with a fixed number of traces, typically single-trace operators.  
It is possible to go beyond the planar limit and do a double expansion in $\lambda$ and $\frac{1}{N}$.
We refer to the lectures of G. Semenoff for a more detailed discussion of the 't Hooft limit and the large-N expansion.

From the action \eqref{eq: SYM-action}, we can derive the propagators. For example, the scalar propagator reads
\begin{equation}
\langle [\phi_i]_{ab}(x) [\phi_j]_{b'a'}(y) \rangle = \delta_{ij}\delta_{aa'}\delta_{bb'} \frac{g_\YM^2}{8\pi^2}\frac{1}{(x-y)^2}\eqndot
  \label{eq:flat-scal-propagator unregularised}
\end{equation}

\subsubsection{Symmetries}
\label{sec: symmetries}
$\mathcal{N}=4$ SYM theory exhibits an exceptional amount of symmetry. In its presentation, we follow the notation in \cite{Beisert:2004ry}. The simplest of its symmetries is given by
Poincar{\' e} symmetry, consisting of the six Lorentz transformations 
 $M_{\mu\nu}$
and the four translations $P_\mu$. 
When treating fermions and dealing with supersymmetry, it is advantageous to exploit the decomposition of the Lorentz group $SO(1,3)\simeq SU(2)_L\times SU(2)_R$. 
The generators of the Lorentz group then are $L^\alpha{}_\beta$ and $\dot{L}^{\dot\alpha}{}_{\dot\beta}$.
Moreover, the momentum generator can be written in terms of spinor indices as $P_{\alpha\dot\alpha}=P_\mu \sigma^{\mu}_{\alpha\dot\alpha}$, where $\sigma^\mu=(\idm,\sigma^1,\sigma^2,\sigma^3)$ with $\sigma^i$ being the Pauli matrices.
The commutation relations of the Lorentz generators among themselves and with the momentum generator $P_{\alpha\dot\alpha}$ follow some general rules. 
For any generator $J$,
\begin{equation}
\label{eq: Lorentz transformations}
 \begin{aligned}
\comm{L^\alpha{}_\beta}{J_\gamma}=\delta_\gamma^\alpha J_\beta-\tfrac{1}{2}\delta_\beta^\alpha J_\gamma\eqncom
\qquad
\comm{L^\alpha{}_\beta}{J^\gamma}=-\delta^\gamma_\beta J^\alpha+\tfrac{1}{2}\delta_\beta^\alpha J^\gamma\eqncom
\\
\comm{\dot{L}^{\dot\alpha}{}_{\dot\beta}}{J_{\dot\gamma}}=\delta_{\dot\gamma}^{\dot\alpha} J_{\dot\beta}-\tfrac{1}{2}\delta_{\dot\beta}^{\dot\alpha} J_{\dot\gamma}\eqncom
\qquad
\comm{\dot{L}^{\dot\alpha}{}_{\dot\beta}}{J^{\dot\gamma}}=-\delta^{\dot\gamma}_{\dot\beta} J^{\dot\alpha}+\tfrac{1}{2}\delta_{\dot\beta}^{\dot\alpha} J^{\dot\gamma}\eqncom
 \end{aligned}
\end{equation}
which is understood to be applied for each index.
\begin{exc}
 Using these rules, write down the explicit commutation relations $\comm{L^{\alpha}{}_{\beta}}{L^{\gamma}{}_{\delta}}$.
\end{exc}
\noindent
Two translations commute
\begin{equation}
 \comm{P_{\alpha\dot\alpha}}{P_{\beta\dot\beta}}=0\eqndot
\end{equation}

In addition, $\mathcal{N}=4$ SYM theory is conformally invariant.
At the classical level, this follows from the absence of masses and dimensionfull couplings in the action \eqref{eq: SYM-action}. The fact that this symmetry is preserved at the quantum level is however non-trivial~\cite{Sohnius:1981sn,Howe:1983sr,Brink:1982wv}.
Conformal symmetry in particular implies the invariance under scale transformations, generated by the dilatation operator $D$, and so-called special conformal transformations, generated by $K^{\alpha\dot\alpha}=K_\mu(\sigma^\mu)^{\alpha\dot\alpha}$. They satisfy the commutation relations
\begin{equation}
\begin{aligned}
 &\comm{D}{P_{\alpha\dot\alpha}}=P_{\alpha\dot\alpha}\eqncom
 \qquad
 \comm{D}{L^\alpha{}_\beta}=\comm{D}{\dot{L}^{\dot\alpha}{}_{\dot\beta}}=0\eqncom
 \qquad
 \comm{D}{K^{\alpha\dot\alpha}}=-K^{\alpha\dot\alpha}\eqncom\\
 &\comm{K^{\alpha\dot\alpha}}{P_{\beta\dot\beta}}
 =\delta^{\dot\alpha}_{\dot\beta}L^{\alpha}{}_\beta
 +\delta^{\alpha}_{\beta}L^{\dot\alpha}{}_{\dot\beta}
 +\delta^{\alpha}_{\beta}\delta^{\dot\alpha}_{\dot\beta}D 
 \eqndot
 \end{aligned}
\end{equation}
The dilatations and special conformal transformations combine with the Poincar\'{e} transformations to form the conformal group $SO(2,4)\simeq SU(2,2)$.

Furthermore, Poincar\'{e} symmetry can be enhanced by supersymmetry, of which $\mathcal{N}=4$ SYM theory has the maximal amount permitted in a theory without gravity.
The supercharges $Q_{\alpha}^{A}$ and $\dot{Q}_{\dot\alpha A}$ have the following non-vanishing anticommutation relations among themselves:
\begin{equation}
\label{eq: QQ P}
 \acomm{\dot{Q}_{\dot\alpha A}}{{Q}_{\alpha}^B}=\delta_A^B P_{\alpha\dot\alpha}\eqncom
\end{equation}
while their behaviour under Lorentz transformations is determined via \eqref{eq: Lorentz transformations}.
Maximal supersymmetry implies a bosonic R-symmetry with symmetry group $SU(4)\simeq SO(6)$. The behaviour of a general generator $J$ under R-symmetry transformations is determined by the following rule in analogy to \eqref{eq: Lorentz transformations}:
\begin{equation}
\comm{R^A{}_B}{J_C}=\delta_C^A J_B-\tfrac{1}{4}\delta_B^A J_C\eqncom
\qquad
\comm{R^A{}_B}{J^C}=-\delta^C_B J^A+\tfrac{1}{4}\delta_B^A J^C\eqndot
\end{equation}

Finally, supersymmetry and conformal symmetry combine to superconformal symmetry with the superconformal charges $S^{\alpha}_A$ and $\dot{S}^{\dot\alpha A}$.
The additional non-vanishing (anti)commutation relations are
\begin{equation}
\begin{aligned}
 &\comm{D}{Q_{\alpha}^{A}}=\tfrac{1}{2} Q_{\alpha}^{A}\eqncom
 \quad
 \comm{D}{\dot{Q}_{\dot\alpha  A}}=+\tfrac{1}{2} \dot{Q}_{\dot\alpha  A}
  \eqncom
  \quad
\comm{D}{S^{\alpha}_{ A}}=-\tfrac{1}{2} S^{\alpha}_{ A}\eqncom
 \quad
 \comm{D}{\dot{S}^{\dot\alpha A}}=-\tfrac{1}{2} \dot{S}^{\dot\alpha A}\eqncom
 \\
 &\acomm{\dot{S}^{\dot\alpha A}}{S^{\alpha}_{B}} =\delta^{A}_{B}K^{\alpha\dot\alpha}\eqncom
 \qquad 
 \comm{K^{\alpha\dot\alpha}}{Q_{\beta}^{A}} =\delta^{\alpha}_{\beta}\dot{S}^{\dot\alpha A}\eqncom
 \qquad
 \comm{K^{\alpha\dot\alpha}}{\dot{Q}_{\dot\beta A}} =\delta_{\dot\beta}^{\dot\alpha}S^{\alpha}_{ A}\eqncom
 \\
 &\acomm{S^{\alpha}_A}{Q_{\beta}^B}
 =\delta_{A}^{B}L^{\alpha}{}_\beta
 +\delta^{\alpha}_{\beta}R^{B}{}_{A}
 +\frac{1}{2}\delta^{\alpha}_{\beta}\delta^{A}_{B}D\eqncom
  \qquad
\comm{S^{\alpha}_{A}}{P_{\beta\dot\beta}} =\delta^{\alpha}_{\beta}\dot{Q}_{\dot\beta A}\eqncom
 \\
 &\acomm{\dot{S}^{\dot\alpha A}}{\dot{Q}_{\dot\beta B}}
 =\delta^{A}_{B}\dot{L}^{\dot\alpha}{}_{\dot\beta}
 -\delta^{\dot\alpha}_{\dot\beta}R^{A}{}_{B}
 +\frac{1}{2}\delta^{\dot\alpha}_{\dot\beta}\delta^{A}_{B}D\eqncom
 \qquad
 \comm{\dot{S}^{\dot\alpha A}}{P_{\beta\dot\beta}} =\delta_{\dot\beta}^{\dot\alpha}Q_{\beta}^{ A}\eqndot
 \label{eq: superconformal}
 \end{aligned}
\end{equation}
Together, they generate the superconformal group $PSU(2,2|4)$.
For the action of $PSU(2,2|4)$ on composite operators, see e.g.\ \cite{Beisert:2004ry}. 

Composite operators $\mathcal{O}_i$ that are primary states of  $PSU(2,2|4)$ can be characterised via the charges $[\Delta, j_L, j_R, r_1, r_2,r_3]$.
Primary means that the operators are annihilated by all lowering operators $\{K^{\alpha\dot\alpha},\, S^{\alpha}_A,\, \dot{S}^{\dot\alpha A},\, L^{\alpha}{}_\beta (\alpha<\beta),\, \dot{L}^{\dot\alpha}{}_{\dot\beta} (\dot\alpha<\dot\beta),\, R^{A}{}_B (A<B)\}$.
All other operators, called descendents, can then be obtained by acting on the primaries with the raising operators
$\{P_{\alpha\dot\alpha},\, Q_{\alpha}^A,\, \dot{Q}_{\dot\alpha A},\, L^{\alpha}{}_\beta (\alpha>\beta),\, \dot{L}^{\dot\alpha}{}_{\dot\beta} (\dot\alpha>\dot\beta),\, R^{A}{}_B (A>B)\}$.
The conformal dimension of the operator, $\Delta$, is measured by the dilatation operator $D$
 and defines the behaviour of the operator under a scale transformation
\begin{equation}\label{dilatation}
x\rightarrow x'=\lambda \, x\eqncom \hspace{0.5cm} \mathcal{O}_i(x) \rightarrow \mathcal{O}_i'(x)= \lambda^{\Delta_i}\mathcal{O}_i(\lambda x)\eqndot
\end{equation}
It will play a particular role in the following, as it can receive quantum corrections.
The other charges are the left and right spin $j_L$ and $j_R$ as well as the three charges $r_1$, $r_2$ and $r_3$ characterising the $SU(4)$ representation in which the operator transforms.

A particular class of primary operators are also annihilated by some of the supercharges $Q_{\alpha}^A$ and $\dot{Q}_{\dot\alpha A}$, which are raising operators. Such primary operators are called BPS operators. 
From the anticommutation relations \eqref{eq: superconformal}, it follows that their scaling dimensions are related to their spin and R-charge and hence protected from quantum corrections.

\subsubsection{Correlation functions}

In conformal field theories, conformal symmetry greatly restricts the form correlation functions can take. For instance, one-point functions of composite operators $\cO_i$ have to be constant  by conformal symmetry and are normally taken to vanish. More generally, all correlation functions are fixed in terms of the so-called conformal data $(\Delta,\lambda)$. The $\Delta$'s are the conformal dimensions of the operators and the $\lambda$'s are called structure constants and describe three-point functions. 

More precisely, the space-time dependence of two-point functions is completely fixed by the scaling dimensions of the operators
\begin{equation}
\label{twopoint}
  \langle \mathcal{O}_i(x) \bar{\mathcal{O}}_j(y) \rangle = \frac{M_{ij}}{|x-y|^{\Delta_i + \Delta_j}}\eqncom
\end{equation}
where $M_{ij} = 0$ for $\Delta_i \neq \Delta_j$.
Moreover, conformal symmetry also fixes the three-point function up to the structure constant $\lambda_{i j k}$, which appears in the operator product expansion (OPE):
\begin{align}\label{eq:GeneralOPE}
\mathcal{O}_i(x) \mathcal{O}_j(y) = \frac{M_{ij}}{|x-y|^{\Delta_{i}+\Delta_{j}}} + \sum_k \frac{\lambda_{i j}{}^{k}} {|x-y|^{\Delta_{i}+\Delta_{j}-\Delta_k} }\,C(x-y,\partial_y)\mathcal{O}_k(y)\eqncom
\end{align}
where the sum over $k$ runs over conformal primary operators and the differential operator $C$ in \eqref{eq:GeneralOPE} accounts for the presence of conformal descendants. The indices on $\lambda$ can be raised and lowered with the matrix $M$. The normalisation of $C$ is such that 
$C(x-y,\partial_y)=1+O(x-y)$.
The scaling dimensions $\Delta_i$ and the structure constants $\lambda_{i j k }$, completely determine all four- and higher-point functions via repeated use of the OPE \eqref{eq:GeneralOPE}. Note that starting from four-point functions, a non-trivial dependence on conformal cross-ratios can occur.

\subsection{Two-point functions and the spectral problem\label{spectrum}}

Let us now calculate the two-point functions.
For simplicity, we will in the following restrict ourselves to the leading large-$N$, planar limit. 
To keep the computation manageable, we further restrict to operators made only of the six scalar fields, the so-called 
$SO(6)$ sector.%
\footnote{Note that this sector is not closed beyond one-loop order.}
In more detail, let 
$I = \{i_1,i_2,\ldots, i_L\}$, with $i_n = 1,\ldots,6$. We then define our un-renormalised operators by
\begin{equation}
  \Obare_I = \tr[\phi_{i_1}\phi_{i_2}\cdots\phi_{i_L}]\eqndot
\end{equation}
It is clear that $\Delta^{(0)} = L$. As correlation functions between
single- and multi-trace operators are suppressed by powers of $\frac{1}{N}$, it is consistent to only consider
single-trace operators.  Since our operators do not contain derivatives, they are conformal primaries in the
sense that $[K_\mu,\Obare_I(0)] = 0$. It is obvious from the propagator \eqref{eq:flat-scal-propagator unregularised}
that the two-point functions 
take  the predicted form~(\ref{twopoint}) at tree level. More precisely, we can write 
\begin{equation}
       \langle \Obare_I(x) \Obarebar_J(y) \rangle_{\mbox{\tiny tree}} \propto \frac{1}{|x-y|^{2\Delta^{(0)}}}
          \delta_{IJ}\eqncom 
\end{equation}
where $\Delta^{(0)}$ is the common classical scaling dimension of $\Obare_I$ and $\Obare_J$, obtainable by
standard power counting. Due to the cyclic invariance of the trace, we identify indices that are cyclic permutations of each other. The bar denotes hermitian conjugation, which in the present case of real scalars only inverts the order of the fields in the trace.

At the quantum level, one observes the phenomenon of operator mixing, meaning that the two-point function between single-trace operators is no longer proportional to a delta-function. Furthermore, wave-function renormalisation is needed in order to render the correlation functions finite and a  regularisation method has to be chosen. In the following, we will make use of dimensional regularisation, i.e.\
\begin{equation}
  S = \frac{2}{g_\YM^2}\int\de^4 x \,\mathcal{L} \to 
  S_\peps = \frac{2}{(g_\YM\mu^\peps)^2}\int\de^{4-2\peps} x \,\mathcal{L}\eqncom
\end{equation}
where $\mu$ is a parameter with the dimension of mass. With this choice of regulator, dimensional analysis shows that
the full two-point function takes the form
\begin{equation}
  \langle \Obare_I(x) \Obarebar_J(y) \rangle_\peps = \sum_{n=0}^\infty (g\mu^\peps)^{2(\Delta^{(0)}+n)}
       \frac{\Mbare_{IJ}^{(n)}(\peps)}{|x-y|^{2\Delta^{(0)}-2\peps(\Delta^{(0)}+n)}}\eqncom
  \label{eq:Obare-two-point}
\end{equation}
where we defined the effective planar loop coupling 
\begin{equation}
  g^2 = \frac{g_\YM^2 N}{16\pi^2}\eqndot
\end{equation}
In general, the $\Mbare_{IJ}^{(n)}(\peps)$ will have poles at $\peps = 0$, so one cannot simply take the
$\peps \to 0$ limit of \eqref{eq:Obare-two-point}. Usually, such divergencies are dealt with by adding counterterms
to the action. For $\mathcal{N}=4$ SYM theory, it is not necessary to introduce such terms; instead one can render the correlation
functions finite by a `rescaling' of the operators alone.\footnote{Usually, this fact is expressed in the abbreviated form
`$\mathcal{N}=4$ SYM theory is finite'.} 
We thus introduce renormalised operators by
\begin{equation}
  \Oren_I = \renZ_{IJ}(g_\YM,\peps) \Obare_J\eqncom
\end{equation} 
where $\renZ_{IJ}(g_\YM,\peps)$ is some numerical matrix with poles at $\peps = 0$ such that the correlation functions
of $\Oren_{I}$ are finite. 

In perturbation theory, if there are no conformal anomalies, the two-point function of the renormalised operators takes
the form \eqref{twopoint}, but with $\Delta_I$ a power series in $g$:
\begin{align}
  \Delta_I = \sum_{n=0}^\infty g^{2n} \Delta_I^{(n)}\eqndot
  \label{eq:Delta-expansion}
\end{align}
For historical reasons, the correction $\Delta - \Delta^{(0)}$ is called the anomalous dimension, even though there
is nothing anomalous about it. Looking at \eqref{eq:Obare-two-point}, what must happen is that the $\log (x-y)^2$ terms resulting
from the expansion of the summand in $\peps$ must exponentiate to form $[(x-y)^2]^\Delta$. To see the exact mechanism,
let us consider the simplified situation with only one operator. We first rewrite the bare two-point function as
\begin{align}
  \langle \Obare(x) \Obarebar(y) \rangle_\peps 
    &= \frac{(g\mu^\peps)^{2\Delta^{(0)}}\Mbare^{(0)}}{[(x-y)^2]^{(1-\peps)\Delta^{(0)}}}
      \left(1+
  \sum_{n=1}^\infty (g^2[\mu^2(x-y)^2]^\peps)^n
       \frac{\Mbare^{(n)}}{\Mbare^{(0)}}\right)\nonumber\\
    &= \frac{(g\mu^\peps)^{2\Delta^{(0)}}\Mbare^{(0)}}{[(x-y)^2]^{(1-\peps)\Delta^{(0)}}}
      \exp\left(\sum_{n=1}^\infty (g^2[\mu^2(x-y)^2]^\peps)^n \logMbare^{(n)}\right)\eqndot
\end{align}
At the second equality, we take the formal logarithm of the series. It is easy to see that the $\logMbare^{(n)}$ are expressible 
as polynomials in $\Mbare^{(n)}/\Mbare^{(0)}$:
\begin{equation}
  \logMbare^{(1)} = \frac{\Mbare^{(1)}}{\Mbare^{(0)}}\eqncom\qquad 
  \logMbare^{(2)} = \frac{\Mbare^{(2)}}{\Mbare^{(0)}}-\frac{1}{2} \left(\frac{\Mbare^{(1)}}{\Mbare^{(0)}}\right)^2\eqncom \qquad 
  \text{etc.}
\end{equation}
If we also write $\renZ$ as a exponential,
\begin{equation}
  \renZ = \exp\left(\sum_{n=1}^\infty g^{2n} \renZ^{(n)}\right)\eqncom
\end{equation}
the renormalised two-point function is\footnote{In this simple example with only one operator,
  we can clearly take $\renZ^{(n)}$ to be real without loss of generality.}
\begin{align}
  \langle \Oren(x) \Orenbar(y) \rangle_\peps 
    &= \renZ \langle \Obare(x) \Obarebar(y) \rangle_\peps \renZ^\dagger\nonumber\\
    &\propto \exp\left(\sum_{n=1}^\infty g^{2n}\left( [\mu^2(x-y)^2]^{\peps n} \logMbare^{(n)} + 2\renZ^{(n)}\right) \right)\eqndot
  \label{eq:renormalised-two-point-exponential}
\end{align}
For this to be finite, we must be able to cancel all divergences with some appropriate choice of $\renZ$. If 
$\logMbare^{(n)}$ had poles of higher degree than one, we would have divergent terms with a dependence on $(x-y)^2$,
which could clearly not be cancelled. We conclude that we can expand $\logMbare^{(n)}$ as
\begin{equation}
  \logMbare^{(n)} = -\frac{\Delta^{(n)}}{n\peps} + \logMbare^{(n),\text{fin}} + O(\peps)\eqndot
  \label{eq:hatM-expansion}
\end{equation}
The choice of the name of the $1/\peps$ coefficient will become clear in a moment. 

From \eqref{eq:renormalised-two-point-exponential} and \eqref{eq:hatM-expansion}, it is obvious that 
a consistent choice of $\renZ$ is
\begin{equation}
  \renZ^{(n)} = \frac{\Delta^{(n)}}{2n\peps}\eqncom
\end{equation}
which only cancels the pole (minimal subtraction).
With this, we can now take the limit $\peps \to 0$ and find
\begin{align}
    &\langle \Oren(x) \Oren(y) \rangle_{\peps = 0}\nonumber\\
      &\quad= \frac{g^{2\Delta^{(0)}}\Mbare^{(0)}(\peps=0)}{[(x-y)^2]^{\Delta^{(0)}}}
        \exp\left(\sum_{n=1}^\infty g^{2n}\left(\logMbare^{(n),\text{fin}}
                                              - \Delta^{(n)}\log \mu^2 (x-y)^2 \right)\right)\nonumber\\
      &\quad= \frac{(g\mu)^{2\Delta^{(0)}}\Mbare^{(0)}(\peps=0)}{[\mu^2(x-y)^2]^{\Delta}}
        \exp\left(\sum_{n=1}^\infty g^{2n}\logMbare^{(n),\text{fin}}\right)\eqncom
\end{align}
with $\Delta$ as defined in \eqref{eq:Delta-expansion}. We see that the divergences of the 
bare correlation functions have transformed into corrections to the scaling dimensions of the renormalised operators.
Note that the mass scale $\mu$, which one could fear would spoil the conformality of the theory, ends up as a harmless
overall normalisation constant. For this reason, in the literature it is common to simply set $\mu = 1$.
With more than one operator, the above considerations still apply, with the added technical complication that $\logMbare^{(n)}$
and $\renZ^{(n)}$ are matrices that do not necessarily commute.

\begin{exc} 
Repeat, up to order $g^4$, the above analysis in the general case with several operators.
For simplicity, assume that $\Mbare^{(0)}$ is proportional to the identity, and that
the bare two-point function looks like
\begin{multline}
 \langle \Obare(x) \Obarebar(y) \rangle_\peps \propto
      \exp\biggl(g^2[\mu^2(x-y)^2]^\peps\left(-\frac{D^{(1)}}{\peps}
                    + \logMbare^{(1),\text{fin}} + O(\peps)\right)\\
                -g^4[\mu^2(x-y)^2]^{2\peps}\left(\frac{D^{(2)}}{2\peps}+O(\peps^0)\right)
                +O(g^6) \biggr)\eqncom
\end{multline}
where $D^{(1)}$ (but not $D^{(2)}$) is diagonal. Partial answer: Let $R$ be a Hermitian
matrix such that $D^{(2)}+i[R,D^{(1)}]$ is diagonal, and set
\begin{equation}
  \renZ = \exp\biggl(g^2\left[\frac{1}{2\peps}D^{(1)}+ iR-\frac{1}{2}\logMbare^{(1),\text{fin}}\right]
    + g^4 \left[\frac{1}{4\peps}(D^{(2)}+i [R,D^{(1)}]) + O(\peps)\right]\biggl)\eqndot
  \label{eq:two-loop-Z}
\end{equation}
Then, the renormalised two-point function is finite, and the anomalous dimensions are the entries
of the diagonal matrix
\begin{equation}
  g^2 D^{(1)} + g^4(D^{(2)}+i [R,D^{(1)}])\eqndot
\end{equation}
\end{exc}

The  higher loop calculations are most conveniently carried out in momentum space, where the 
contraction rules read
\begin{equation}
  \langle [\phi_i]_{ab} [\phi_j]_{b'a'} \rangle = \frac{(g_\YM\mu^\peps)^2}{2}\frac{\delta_{ij}\delta_{aa'}\delta_{bb'}}{p^2}\eqncom
  \hspace{0.5cm}
 \langle [A_\mu]_{ab} [A_\nu]_{b'a'} \rangle = \frac{(g_\YM\mu^\peps)^2}{2}\frac{\delta_{\mu\nu}\delta_{aa'}\delta_{bb'}}{p^2}\eqndot
\end{equation}
The transition from momentum space to configuration space is encoded in the formula
\begin{equation}
  \int \frac{\de^{4-2\peps} p}{(2\pi)^{4-2\peps}}\frac{\e^{ip\cdot x}}{[p^2]^s} 
    = \frac{\Gamma(2-\peps-s)}{4^s\pi^{2-\peps}\Gamma(s)}\frac{1}{[x^2]^{2-\peps-s}}\eqndot
  \label{eq:fourier-transform}
\end{equation}
We denote by
\begin{equation}
  K(x,y) = \frac{(g_\YM\mu^\peps)^2}{2} \int \frac{\de^{4-2\peps} p}{(2\pi)^{4-2\peps}}\frac{\e^{ip\cdot (x-y)}}{p^2} 
  = \frac{(g_\YM\mu^\peps)^2}{2}\frac{\Gamma(1-\peps)}{4\pi^{2-\peps}}\frac{1}{[(x-y)^2]^{1-\peps}}\eqncom
  \label{eq:flat-scal-propagator}
\end{equation}
the scalar propagator in $(4-2\peps)$-dimensional position space.

This results in the following more specific form of the two-point function of bare operators:
\begin{equation}
       \langle \Obare_I(x) \Obarebar_J(y) \rangle_\peps = \sqrt{c_Ic_J} 
         N^{\Delta^{(0)}} K(x,y)^{\Delta^{(0)}}
          \biggl(\delta_{IJ} 
          +\sum_{n=1}^\infty g^{2n}\Mbare_{IJ}^{(n)}(\peps)[\mu^2|x-y|^2]^{n\peps}\biggr)\cdot
\end{equation}
For convenience, we have pulled out the tree-level two-point function. 
The normalisation constants $c_I$  are easily seen to be
the number of cyclic permutations that leave  $I = \{i_1,i_2,\ldots, i_L\}$ 
invariant. For example,
\begin{equation}
  c_{\{5,5,5,5\}} = 4\eqncom\quad c_{\{1,5,5,5\}} = 1\eqncom\quad c_{\{1,5,5,1,5,5\}} = 2 \eqndot
\end{equation}
As we will explicitly show below, the divergence at one loop is a simple $1/\peps$ pole. We can thus write
\begin{equation}
  \Mbare^{(1)} = -\frac{1}{\peps}D^{(1)} + \Mbare^{(1),\text{fin}} + O(\peps)\eqncom
\end{equation}
with $D^{(1)}$ and $ \Mbare^{(1),\text{fin}}$ independent of $\peps$. We now choose our renormalisation scheme as follows:
\begin{equation}
  \Oren_I = \renZ_{IJ}(g_\YM,\peps) \frac{\Obare_J}{\sqrt{c_J}}\eqncom
\end{equation} 
with 
\begin{equation} \label{renZ}
  \renZ = 1+\frac{g^2}{2\peps}D^{(1)}-\frac{g^2}{2}\Mbare^{(1),\text{fin}}\eqndot
\end{equation}
The motivation for this particular choice for the finite part is two-fold. First, as will become evident
below, it leaves us with only one matrix to diagonalise and secondly, as will become clear
in section~\ref{one-loop-one-point} it implies a 
convenient normalisation of the renormalised operators.
The choice of $\renZ$ in~(\ref{renZ}) results in the following 
expression for the renormalised two-point function:
\begin{equation} \label{twopointren}
       \langle \Oren_I(x) \Orenbar_J(y) \rangle_{\peps=0} = 
         N^{\Delta^{(0)}} K(x,y)^{\Delta^{(0)}}
          \biggl(\delta_{IJ}
            -g^2D^{(1)}_{IJ}\log(\mu^2|x-y|^2)+O(g^4)\biggr)\eqndot
\end{equation}
It is clear from (\ref{twopointren}) that, in order to determine the good conformal operators at the one-loop level
and their corresponding conformal dimensions, one has to diagonalise the matrix $D^{(1)}_{IJ}$ which is denoted as diagonalising
the  dilatation operator. More precisely, the eigenvectors of $D^{(1)}$ are the good conformal operators and the corresponding eigenvalues are the associated conformal dimensions at one-loop order.

Said more formally, we should really set
\begin{equation}
  \renZ = U\left(1+\frac{g^2}{2\peps}D^{(1)}-\frac{g^2}{2}\Mbare^{(1),\text{fin}}\right)\eqncom
\end{equation}
where $U$ is a unitary matrix such that $UD^{(1)}U^\dagger$ is diagonal. Then, the renormalised two-point function
takes the proper form \eqref{twopoint}, with
\begin{equation}
  \Delta^{(1)}_I = [UD^{(1)}U^\dagger]_{II}\eqndot
\end{equation}
Let us remark that it is necessary to know the two-point function at order $g^2$
to determine the $g^0$ piece of $\renZ$ (i.e.~$U$). This happens because we are really doing degenerate perturbation
theory in the sense that several operators have the same value of $\Delta^{(0)}$. As one can explicitly see from
\eqref{eq:two-loop-Z}, this pattern continues at higher loop order (i.e.~$\renZ$ at one loop depends on the
two-loop correlation function).
The remainder of this section will be dedicated to explicitly determining $D^{(1)}$ (and $\Mbare^{(1),\text{fin}}$) by a 
Feynman diagram calculation.

\begin{figure}[t]
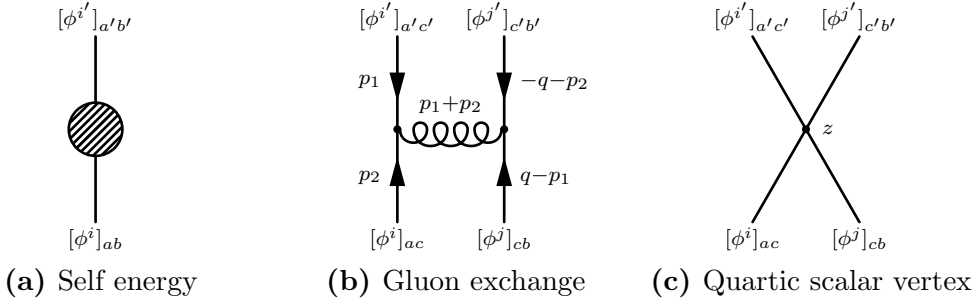

\fmfstraight
 \begin{subfigure}{0.3\textwidth}
 \centering
  \fmfframe(10,10)(10,10){%
 \begin{fmfchar*}(40,70)
 \fmftop{o1}
 \fmfbottom{i1}
 \fmf{plain}{i1,v}
 \fmf{plain}{o1,v}
 \fmfblob{20}{v}
 \fmffreeze
 \fmfposition
 \fmfiv{label=$\scriptstyle [\phi_i]_{ab}$,l.dist=2}{vloc(__i1)}
 \fmfiv{label=$\scriptstyle [\phi_{i'}]_{b'a'}$,l.dist=2}{vloc(__o1)}
 \end{fmfchar*}
 }
  \caption{Self energy}
  \label{subfig: self energy}
 \end{subfigure}
   \begin{subfigure}{0.3\textwidth}
 \centering
  \fmfframe(10,10)(10,10){%
 \begin{fmfchar*}(40,70)
 \fmftop{o1,o2}
 \fmfbottom{i1,i2}
 \fmf{electron,label=$\scriptstyle p_2$,l.s=left}{i1,v1}
 \fmf{electron,label=$\scriptstyle p_1$}{o1,v1}
 \fmf{electron,label=$\scriptstyle q-p_1$}{i2,v2}
 \fmf{electron,label=$\scriptstyle -q-p_2$,l.s=left}{o2,v2}
 \fmf{gluon,label=$\scriptstyle p_1+p_2$,l.s=left,tension=0}{v1,v2}
 \fmfv{decor.shape=circle,decor.filled=full,
decor.size=thick}{v1}
 \fmfv{decor.shape=circle,decor.filled=full,
decor.size=thick}{v2}
 \fmffreeze
 \fmfposition
 \fmfiv{label=$\scriptstyle [\phi_i]_{ac}$,l.a=-90,l.dist=2}{vloc(__i1)}
 \fmfiv{label=$\scriptstyle [\phi_j]_{cb}$,l.a=-90,l.dist=2}{vloc(__i2)}
 \fmfiv{label=$\scriptstyle [\phi_{i'}]_{c'a'}$,l.a=90,l.dist=2}{vloc(__o1)}
 \fmfiv{label=$\scriptstyle [\phi_{j'}]_{b'c'}$,l.a=90,l.dist=2}{vloc(__o2)}
 \end{fmfchar*}
 }
  \caption{Gluon exchange}
  \label{subfig: gluon}
 \end{subfigure}
  \begin{subfigure}{0.3\textwidth}
 \centering
  \fmfframe(10,10)(10,10){%
 \begin{fmfchar*}(40,70)
 \fmftop{o1,o2}
 \fmfbottom{i1,i2}
 \fmf{plain}{i1,v}
 \fmf{plain}{o1,v}
 \fmf{plain}{i2,v}
 \fmf{plain}{o2,v}
 \fmfv{label=$\scriptstyle z$,decor.shape=circle,decor.filled=full,
decor.size=thick}{v}
 \fmffreeze
 \fmfposition
 \fmfiv{label=$\scriptstyle [\phi_i]_{ac}$,l.a=-90,l.dist=2}{vloc(__i1)}
 \fmfiv{label=$\scriptstyle [\phi_j]_{cb}$,l.a=-90,l.dist=2}{vloc(__i2)}
 \fmfiv{label=$\scriptstyle [\phi_{i'}]_{c'a'}$,l.a=90,l.dist=2}{vloc(__o1)}
 \fmfiv{label=$\scriptstyle [\phi_{j'}]_{b'c'}$,l.a=90,l.dist=2}{vloc(__o2)}
 \end{fmfchar*}
 }
  \caption{Quartic scalar vertex}
  \label{subfig: quartic}
 \end{subfigure}
\caption{Interaction part of the Feynman diagrams contributing to the two-point function at one-loop order.
}
 \label{fig: Feynman diagrams}
\end{figure}

In the planar limit, the one-loop corrections to 
the two-point functions consist of three types of diagrams, see figure \ref{fig: Feynman diagrams}. The colour structure is 
completely fixed by the planar limit, so the interesting part is the flavour structure. By $SO(6)$ symmetry,
the self-energy diagram (\subref{subfig: self energy}) must be proportional to $\delta_{ii'}$. Similarly, since the gauge field is not
charged under $SO(6)$, the gauge exchange diagram (\subref{subfig: gluon}) must have the structure $\delta_{ii'}\delta_{jj'}$.
However, the four-point diagram (\subref{subfig: quartic}) allows for non-trivial flavour tensors. This is what leads to operator mixing
at the one-loop level.

It happens that the most interesting diagram is also the easiest to compute, so let us begin by
considering the diagram arising from the four-point interaction $\tr\left([\phi_i,\phi_j][\phi_i,\phi_j]\right)$.
Writing only the two fields of each operator that participate, we find the contribution
\begin{multline}
  \langle [\phi_i\phi_j]_{ab}(x)[\phi_{j'}\phi_{i'}]_{b'a'}(y) \rangle_{\text{(c)}} =
      \frac{2N^2\delta_{aa'}\delta_{bb'}}{(g_\YM\mu^\peps)^2}(2\delta_{ij'}\delta_{ji'}
        -\delta_{ii'}\delta_{jj'}-\delta_{ij}\delta_{i'j'})\\
  \times \int\de^{4-2\peps} z K(x,z)^2 K(z,y)^2 \eqndot
\end{multline}
We now insert the explicit form of the propagator using \eqref{eq:fourier-transform} and \eqref{eq:flat-scal-propagator} 
to get
\begin{multline}
  \cdots = 
      \frac{2N^2\delta_{aa'}\delta_{bb'}}{(g_\YM\mu^\peps)^2}(2\delta_{ij'}\delta_{ji'}
        -\delta_{ii'}\delta_{jj'}-\delta_{ij}\delta_{i'j'})\\
  \times \left[\frac{(g_\YM\mu^\peps)^2}{2}\frac{\Gamma(1-\peps)}{4\pi^{2-\peps}}\right]^4
         \int\de^{4-2\peps} z \frac{1}{[(x-z)^2]^{2-2\peps}[(z-y)^2]^{2-2\peps}} \eqndot
\end{multline}
This is a standard one-loop integral  which can be evaluated using the formula 
(see e.g.~\cite{Smirnov:2006ry})
\begin{equation}
  \int \de^{4-2\peps} p \frac{1}{[p^2]^{\alpha}[(p-q)^2]^\beta} 
    = \pi^{2-\peps}G(\alpha,\beta)\frac{1}{[q^2]^{\alpha+\beta+\peps-2}}\eqndot
  \label{eq:one-loop-master}
\end{equation}
Here, $G(\alpha,\beta)$ denotes the following combination of gamma functions,
\begin{equation}
  G(\alpha,\beta) = \frac{\Gamma(\alpha+\beta+\peps-2)\Gamma(2-\peps-\alpha)\Gamma(2-\peps-\beta)}
                         {\Gamma(\alpha)\Gamma(\beta)\Gamma(4-\alpha-\beta-2\peps)}\eqndot
\end{equation}
We thus find
\begin{align}
  \langle [\phi_i\phi_j]_{ab}(x)[\phi_{j'}\phi_{i'}]_{b'a'}(y) \rangle_{\text{(c)}} &=
      \frac{(g_\YM\mu^\peps)^6 N^2 \Gamma(1-\peps)^4 \delta_{aa'}\delta_{bb'}}
           {2^{11}\pi^{6-3\peps}}\nonumber\\
      &\qquad\qquad\times (2\delta_{ij'}\delta_{ji'}
        -\delta_{ii'}\delta_{jj'}-\delta_{ij}\delta_{i'j'})\frac{G(2-2\peps,2-2\peps)}{[(x-y)^2]^{2-3\peps}}\nonumber\\
      &= g^2 N K(x,y)^2 \delta_{aa'}\delta_{bb'} 
          (2\delta_{ij'}\delta_{ji'} -\delta_{ii'}\delta_{jj'}-\delta_{ij}\delta_{i'j'})\nonumber\\
      &\qquad\qquad\times\left(\frac{1}{\peps} + 1 + \gammaE + \log(\pi|x-y|^2) + O(\peps)\right)
      \eqncom
  \label{eq:four-point-result}
\end{align}
where $\gammaE$ denotes the Euler-Mascheroni constant.

The second type of diagram which involves a pair of legs is the gauge boson exchange. The diagram
is formed using two copies of the vertex $i\tr[(\partial_\mu \phi_i)[A_\mu,\phi_i]]$. Explicitly,
we find the contribution
\begin{multline}
  \langle [\phi_i\phi_j]_{ab}(x)[\phi_{j'}\phi_{i'}]_{b'a'}(y) \rangle_{\text{(b)}} =
   -\frac{4\delta_{ii'}\delta_{jj'}\delta_{aa'}\delta_{bb'}N^2}{(g_\YM\mu^\peps)^4}\\
   \times \int\de^{4-2\peps} z\de^{4-2\peps} w \left(K(x,z)\partial_{z_\mu}K(z,y)-[\partial_{z_\mu}K(x,z)]K(z,y)\right)K(z,w)\\
   \times \left([\partial_{w_\mu}K(x,w)]K(w,y)-K(x,w)\partial_{w_\mu}K(w,y)\right)\eqndot
\end{multline}
In this case, it is simpler to work in momentum space. We thus insert \eqref{eq:flat-scal-propagator} and
integrate over $z,w$ and the momenta fixed by the resulting delta functions to find
\begin{equation}
  \cdots = 
   -\frac{(g_\YM\mu^\peps)^6\delta_{ii'}\delta_{jj'}\delta_{aa'}\delta_{bb'}N^2}{2^3}
   \int\frac{\de^{4-2\peps} q}{(2\pi)^{4-2\peps}}\e^{iq\cdot(x-y)}H(q)\eqncom
\end{equation}
with 
\begin{equation}
  H(q) = \iint
         \frac{(p_1-p_2)\cdot(2q-p_1+p_2)}{V}\eqncom \qquad 
         V = p_1^2p_2^2(p_1-q)^2(p_2+q)^2(p_1+p_2)^2\eqndot
\end{equation}
To save space, we abbreviate
\begin{equation}
  \iint = \int \frac{\de^{4-2\peps} p_1\de^{4-2\peps} p_2}{(2\pi)^{2(4-2\peps)}}\eqncom
\end{equation}
here and in the following.

We now have to perform the integrals over $p_1$ and $p_2$. First, we rewrite the numerator as a linear combination
of the factors in the denominator,
\begin{equation}
  (p_1-p_2)\cdot(2q-p_1+p_2) = -[p_1^2+p_2^2+(p_1-q)^2+(p_2+q)^2]+(p_1+p_2)^2+2q^2\eqndot
\end{equation}
By the evident symmetries of the diagram, we then find 
\begin{equation}
  H = -4H_1 + H_2 + 2q^2 H_3\eqncom
\end{equation}
with 
\begin{equation}
  H_1 = \iint\frac{p_1^2}{V}\eqncom\quad H_2 = \iint\frac{(p_1+p_2)^2}{V}\eqncom\quad H_3 = \iint\frac{1}{V}\eqndot
\end{equation}
The numerator of $H_1$ cancels one of the propagators, allowing us to perform the $p_1$ integral using
\eqref{eq:one-loop-master}. The remaining integral over $p_2$ then also follows from \eqref{eq:one-loop-master}, and
we obtain
\begin{equation}
  H_1 = \frac{1}{(4\pi)^{4-2\peps}}\frac{G(1,1)G(1,1+\peps)}{[q^2]^{2\peps}}\eqndot
\end{equation}
The $p_1$ and $p_2$ integrals decouple for $H_2$, and we immediately find
\begin{equation}
  H_2 = \frac{1}{(4\pi)^{4-2\peps}}\frac{G(1,1)^2}{[q^2]^{2\peps}}\eqndot
\end{equation}

To get a closed expression for the final integral $H_3$ requires an extra trick. 
Here, we use integration by parts, following \cite{Kazakov:1983ns,Smirnov:2006ry}.
We first observe that
\begin{equation}
  0 = \iint(\partial_1\cdot p_1+p_2\cdot\partial_1)\frac{1}{V}
    = -2\peps H_3 + \iint\frac{p_2^2-(p_1+p_2)^2}{p_1^2 V} + \iint\frac{(p_2+q)^2-(p_1+p_2)^2}{(p_1-q)^2V}\eqndot
\end{equation}
We can now isolate $H_3$ and evaluate the remaining integrals by successive use of \eqref{eq:one-loop-master},
with the result
\begin{equation}
  H_3 = \frac{1}{\peps}\iint\frac{p_2^2-(p_1+p_2)^2}{p_1^2 V}
      = \frac{1}{\peps}\frac{1}{(4\pi)^{4-2\peps}}\frac{G(1,1)[G(2,1+\peps)-G(2,1)]}
                                                             {[q^2]^{1+2\peps}}\eqndot
\end{equation}
After the transformation back to real space, the final result for the diagram is 
\begin{multline}
  \langle [\phi_i\phi_j]_{ab}(x)[\phi_{j'}\phi_{i'}]_{b'a'}(y) \rangle_{\text{(b)}} =\\
   g^2 N K(x,y)^2 \delta_{ii'}\delta_{jj'}\delta_{aa'}\delta_{bb'} \left(\frac{1}{\peps}
   + 3 + \gammaE + \log(\pi|x-y|^2) + O(\peps)\right)\eqndot
  \label{eq:gauge-exchange-result}
\end{multline}
Let us remark that $H_3$ is finite in four dimensions, in fact
\begin{equation}
  H_3 = \frac{3\zeta(3)}{2^7\pi^4}+O(\peps)\eqndot
\end{equation}
The Fourier transform yields an additional factor of $\peps$, meaning that the contribution
of $H_3$ to the two-point function is $O(\peps)$. In a one-loop calculation, $H_3$ can thus be dropped.

Finally, we have the  one-loop self-energy correction to the scalar propagator.
The calculation again reduces to an application of \eqref{eq:one-loop-master}.
We omit the details, but see \cite{Erickson:2000af}. The one-loop corrected propagator is
\begin{align}
  &\langle [\phi_i(x)]_{ab} [\phi_j(y)]_{b'a'} \rangle \nonumber\\
    &\quad= \delta_{ij}\delta_{aa'}\delta_{bb}\left[K(x,y)-
       g_\YM^4N \int\frac{\de^{4-2\peps} q}{(2\pi)^{4-2\peps}}
       \e^{iq\cdot(x-y)}\frac{G(1,1)}{(4\pi)^{2-\peps}}\frac{1}{[q^2]^{1+\peps}}
       +O(g_\YM^6)\right]\nonumber\\
  &\quad= \delta_{ij}\delta_{aa'}\delta_{bb}K(x,y)\left[1-2g^2\left(\frac{1}{\peps}
   + 2 + \gammaE + \log(\pi|x-y|^2) + O(\peps) \right)+O(g^4)\right]\eqndot
  \label{eq:self-energy}
\end{align}
The planar two-point function of two single-trace operators now follows by inserting 
the corrections \eqref{eq:four-point-result}, \eqref{eq:gauge-exchange-result} and
\eqref{eq:self-energy} in the tree-level diagram. All told, we find
\begin{multline}
       \langle \Obare_I(x) \Obarebar_J(y) \rangle_{\peps} = \sqrt{c_Ic_J} 
         N^{\Delta^{(0)}} K(x,y)^{\Delta^{(0)}}\\
         \times \biggl[\delta_{IJ}
            -g^2\left(\frac{1}{\peps} + 1 + \gammaE +\log(\pi|x-y|^2)\right)D^{(1)}_{IJ}
            +g^2O(\peps) + O(g^4)\biggr]\eqncom
  \label{eq:SO6-two-point}
\end{multline}
with
\begin{equation}
  D^{(1)}_{IJ} = \frac{1}{\sqrt{c_Ic_J}}\sum_{n=1}^L \left(2-2\permop_{n,n+1}+\traceop_{n,n+1}\right)
  \left(\delta_{i_1,j_1}\delta_{i_2,j_2}\cdots \delta_{i_L,j_L}+
      \text{cyclic perm.}\right)\eqndot
  \label{eq:SO6-D1}
\end{equation} 
This expression requires some explanation. First of all, we identify $L + 1 = 1$.
In the last factor, one should add cyclic permutation of the $j_n$ indices relative
to the $i_n$ indices, e.g.
\begin{multline}
  \delta_{i_1,j_1}\delta_{i_2,j_2}\delta_{i_3,j_3}+
      \text{cyclic permutations}\\
  = \delta_{i_1,j_1}\delta_{i_2,j_2}\delta_{i_3,j_3}
    + \delta_{i_1,j_2}\delta_{i_2,j_3}\delta_{i_3,j_1}
    + \delta_{i_1,j_3}\delta_{i_2,j_1}\delta_{i_3,j_2}\eqndot
\end{multline}
The sum in \eqref{eq:SO6-D1} is understood to be an operator acting
on the Kronecker deltas. Specifically, $\permop_{n,n+1}$ acts on the factors involving 
$i_n$ and $i_{n+1}$ as
\begin{equation}
  \permop_{n,n+1} \left(\cdots \delta_{i_n,j_{m}}\delta_{i_{n+1},j_{m+1}}\cdots\right)
    = \cdots \delta_{i_{n+1},j_{m}}\delta_{i_{n},j_{m+1}}\cdots \eqncom
  \label{eq:P-index-action}
\end{equation}
leaving all other factors invariant. Similarly, the action of $K_{n,n+1}$ is
\begin{equation}
  \traceop_{n,n+1} \left(\cdots \delta_{i_n,j_{m}}\delta_{i_{n+1},j_{m+1}}\cdots\right)
    = \cdots \delta_{i_{n},i_{n+1}}\delta_{j_{m},j_{m+1}}\cdots \eqndot
\end{equation}
In this way, one generates the two non-trivial flavour structures we found
in \eqref{eq:four-point-result}.

There is an important subsector of the $SO(6)$ sector called the $SU(2)$ sector. Here, we
are only allowed to build operator using the two complex scalar fields $X$ and $Y$, defined by
\begin{equation}
  X = \phi_1 + i\phi_4\eqncom\qquad Y = \phi_2 + i\phi_5\eqndot
\end{equation}
The propagators look like 
\begin{equation}
  \langle [X]_{ab}(x) [\bar{X}]_{b'a'}(y) \rangle = 2\delta_{aa'}\delta_{bb'} K(x,y)\eqncom\quad
  \langle [X]_{ab} [X]_{b'a'} \rangle = \langle [\bar{X}]_{ab} [\bar{X}]_{b'a'} \rangle = 0\eqncom
\end{equation}
and similarly for $Y$. Note the extra factor of two. The $SU(2)$ sector is closed to all loop orders,
in contrast to the $SO(6)$ sector. It is easy to deduce the $SU(2)$ dilatation operator from the above
computation. For $S = \{s_1,s_2,\ldots,s_L\}$ with $s_n = \uparrow,\downarrow$, let us define the 
bare operator
\begin{equation}
  \Obare_S = \tr[\phi_{s_1}\cdots \phi_{s_L}]\eqncom  \qquad
    \phi_{\uparrow} = X\eqncom\quad \phi_{\downarrow} = Y\eqndot
\end{equation}
We then find
\begin{multline}
       \langle \Obare_S(x) \Obarebar_{S'}(y) \rangle_{\peps} = \sqrt{c_Sc_{S'}} 
         (2N)^{\Delta^{(0)}} K(x,y)^{\Delta^{(0)}}\\
         \times \biggl[\delta_{SS'}
            -g^2\left(\frac{1}{\peps} + 1 + \gammaE +\log(\pi|x-y|^2)\right)D^{(1)}_{SS'}
            +g^2O(\peps) + O(g^4)\biggr]\eqncom
  \label{eq:SU2-two-point}
\end{multline}
with
\begin{equation}
  D^{(1)}_{SS'} = \frac{2}{\sqrt{c_Sc_{S'}}}\sum_{n=1}^L \left(1-\permop_{n,n+1}\right)
  \left(\delta_{s_1,s'_1}\delta_{s_2,s'_2}\cdots \delta_{s_L,s'_L}+
      \text{cyclic perm.}\right)\eqndot
\end{equation}
Here, $\permop_{n,n+1}$ acts as in \eqref{eq:P-index-action}, but with $s$'s instead of $i$'s. Terms
originating from the $SO(6)$ trace operator $\traceop$ are seen to be proportional to $1+i^2 = 0$.
\begin{exc}
  Check the details of the reduction to the $SU(2)$ sector.
\end{exc}

Since $D^{(1)}_{SS'}$ only involves the permutation operator $\permop$, it is clear that the
two-point function is only non-zero between operators with the same number of $X$ and $Y$ fields.
Using the $SO(6)$ symmetry, it can be shown that this holds to all orders in $g$. When diagonalising
the dilation matrix, one can thus restrict to operators with a fixed number of $X$'s and $Y$'s.
\begin{exc}
Using two $X$'s and two $Y$'s, one can form two bare single-trace operators. Construct the
corresponding (one-loop) renormalised operators, and show that the anomalous dimensions
are $\Delta^{(1)} = 0$ and $\Delta^{(1)} = 12$.
\end{exc}

\subsubsection{Spin chains} 

Now that we have computed the one-loop dilatation operator $D^{(1)}$ in the planar limit, the natural next problem is to find its eigenvectors and eigenvalues. This is called the spectral problem.
Let us consider operators of length $L$, so that their classical conformal dimension is $\Delta^{(0)}=L$. These operators form a vector space on which the dilatation operator acts. A general operator in this space is of the form
\begin{align}
\cO = \Psi^{S} \Obare_{S}\eqncom
\label{eq:O-from-Psi}
\end{align}
where the coefficient $\Psi^{S}$ is invariant under cyclic permutations,
\begin{equation}
  \Psi^{\{s_1,s_2,\ldots,s_L\}} = \Psi^{\{s_L,s_1,s_2,\ldots,s_{L-1}\}} \eqndot
\end{equation}
Now $\Psi$ can be seen as a vector in $\mathbb{C}^{2L}$ (more precisely the cyclically invariant subspace of $\mathbb{C}^{2L}$) since each index $s_i=\uparrow,\downarrow$ takes two values. 
In this language, $D^{(1)}_{SS^\prime}$ actually defines an operator $H: \mathbb{C}^{2L}\rightarrow \mathbb{C}^{2L}$.
To see this precisely, we note that the two-point function of two operators of the form \eqref{eq:O-from-Psi} is
\begin{multline}
  \langle\cO_1(x)\bar{\cO}_2(y)\rangle_\peps = L (2N)^{\Delta^{(0)}}K(x,y)^{\Delta^{(0)}} \\
  \times \langle \Psi_2 | 1
            -g^2\left(\frac{1}{\peps} + 1 + \gammaE +\log(\pi|x-y|^2)\right)H
            +g^2O(\peps) + O(g^4) | \Psi_1 \rangle \eqndot
  \label{eq:SU2-two-point-spin-chain}
\end{multline}
Here, the inner product is the usual
\begin{equation}
  \langle\Psi_2 | \Psi_1 \rangle = (\Psi^S_2)^*\Psi^S_1\eqncom
\end{equation}
and $H$ is given by
\begin{align}\label{eq:HfromD}
  H = 2 \sum_{i=1}^L (1 - \mathbb{P}_{i,i+1})\eqncom 
\end{align}
where $\permop_{i,i+1}$ now denotes the operator which permutes two neighboring spins,
\begin{equation}
  (\permop_{i,i+1} \Psi)^{\{s_1,\ldots,s_L\}} = \Psi^{\{s_1,\ldots,s_{i-1},s_{i+1},s_{i},s_{i+2},\ldots,s_L\}}\eqndot
\end{equation}
As a slight abuse of notation, we will also denote $D^{(1)} \equiv H$.
\begin{exc}
  Derive \eqref{eq:SU2-two-point-spin-chain} from \eqref{eq:SU2-two-point}. To get the combinatorial
  factors right, it is important to use that $\Psi_1$ and $\Psi_2$ are invariant under cyclic permutations.
\end{exc}

The fact that $\cO$ is an eigenstate of the dilatation operator then simply translates to $\Psi$ being an eigenstate of $D^{(1)}$. The anomalous dimension of $\cO$ is then the eigenvalue of $\Psi$. It is obvious that the wave function
$\Psi^{\{s_1,\ldots,s_L\}}= \delta_{s_1 \uparrow}\ldots \delta_{s_L \uparrow}$ corresponding to all the spins in the spin chain
state pointing upwards is an eigenstate of $D^{(1)}$  with eigenvalue zero and correspondingly
 the operator ${\mathcal{O}}$  built entirely from $X$-fields is an eigenstate
of the dilatation operator and has anomalous dimension equal to zero.\footnote{Obviously, the argument can be repeated
with $\uparrow$ replaced by $\downarrow$ and $X$ replaced by $Y$.} 
This operator constitutes an example of a BPS operator
and its anomalous dimension vanishes to all loop orders.

The operator $H$ is a sum of terms for which only neighbouring sites interact. For this reason, it is called a nearest-neighbour operator. We can rewrite \eqref{eq:HfromD} in a more familiar way by using the Pauli matrices
\begin{align}
&H = L - 4\sum_{i=1}^L \vec{S}_i \cdot \vec{S}_{i+1} \eqncom
&& \vec{S} = \frac{1}{2}(\sigma_1,\sigma_2,\sigma_3)\eqncom
\end{align}
where $\vec{S}_i$ is the spin operator acting on the $i$'th particle. The operator $H$ is
the Hamiltonian  of the well-known Heisenberg spin chain. This Hamiltonian is integrable, which means that its eigenvalues and eigenvectors can be found by the Bethe ansatz as we will discuss in the next section. 

\subsubsection{Other sectors}
Here, we have derived the one-loop dilatation generator for operators belonging to the $SU(2)$ sector
of $\mathcal{N}=4$ SYM theory but the one-loop dilatation generator can can be determined for any kind of composite operator \cite{Beisert:2003jj}. When diagonalising the dilatation generator, however,  one typically
considers only a sub-sector which  then must be closed under renormalisation. The closed sectors were classified  in~\cite{Beisert:2003jj}. 
Most studied are the closed sectors of rank one of which there exist three.
The simplest of these is the $SU(2)$ sector which
we have treated above. With a step up in complexity one finds the $SL(2)$ sector, which is built from a complex scalar $X$ with arbitrary many covariant derivatives $D_{1\dot1}$ polarised in a single light-like direction. In contrast to the $SU(2)$ sector, the $SL(2)$ sector is non-compact. The third sector of rank one is the so-called $SU(1|1)$ sector which contains a bosonic as well as a fermionic field.
An important phenomenon which occurs starting from two-loop order is that of the dilatation generator
introducing length changing of the operators on which it acts.
The smallest sector that exhibits length-changing is $SU(2|3)$. For a discussion of this sector,
we refer to~\cite{Beisert:2003ys}.
In contrast, the largest sector in which length is preserved is $PSU(1,1|2)$.  This sector is treated
in~\cite{Beisert:2007sk,Zwiebel:2008gr}.

\section{The integrable Heisenberg spin chain in \texorpdfstring{$\mathcal{N}=4$}{N=4} SYM theory}
\label{spinchain}

\subsection{One loop}
A simple way of expressing the fact that the Heisenberg spin chain constitutes an integrable system is by stating that there
exist $L$ local conserved charges $Q_1, Q_2,\ldots, Q_L$ which fulfil
\begin{equation}
[Q_i,Q_j]=0\eqncom \hspace{0.7cm} i,j=1,\ldots,L\eqndot
\end{equation}
The charges can be organised so that $Q_l$ involves interactions between $l$ neighbouring spins. The first charge $Q_1$
can be taken as the total momentum of the spin chain, the conservation of which obviously follows from the translational invariance of the chain.
The second charge $Q_2$ can be taken as the Hamiltonian. The third charge $Q_3$,
which will play a distinguished role in the following, can be chosen as
\begin{equation}
Q_3=\sum_{i=1}^L [H_{i,i+1},H_{i+1,i+2}]\eqncom
\end{equation}
and there exists a certain boosting procedure which makes it possible to construct the remaining higher conserved 
charges~\cite{Tetelman,Sogo}.

Another way of expressing the integrability of the Heisenberg spin chain is by stating that it can be solved by the algebraic Bethe ansatz approach, explained in the lectures by J.L.\ Jacobsen.  In this approach one starts from a reference state
\begin{equation}
 \label{eq: vacuum}
 |0\rangle_L\equiv|\underbrace{\uparrow\dots\uparrow}_L\rangle \eqncom
\end{equation}
with all spins pointing upwards, say, and obtains the other highest-weight 
 eigenstates by acting on the reference state with a number of creation operators $\hat{B}(u)$,
each of which generates an appropriate linear combination of states where one spin-up has been replaced by a spin-down,  
thus lowering the total spin by one unit, i.e.\
\begin{equation}\label{eigenstate}
|\textbf{u}\rangle= \hat{B}(u_1)\ldots \hat{B}(u_M) |0\rangle_L\eqndot
\end{equation}
The creation operators depend on  rapidity variables $u_i$.
In order for $|\textbf{u}\rangle$ to be an eigenstate of the Heisenberg
spin chain, the rapidities have to fulfil the Bethe equations
\begin{align}\label{Betheeqns}
\left(\frac{u_k+\frac{i}{2}}{u_k-\frac{i}{2}}\right)^L =\prod_{\substack{j=1\\j\neq k}}^M \frac{u_k-u_j+i}{u_k-u_j-i}\eqndot
 \end{align}
The corresponding energy eigenvalue is
 \begin{align}\label{Energy}
 E=2\sum_{i=1}^M\frac{1}{u_i^2+\frac{1}{4}}\eqndot
 \end{align} 
 The algebraic Bethe ansatz simultaneously diagonalises all the conserved charges of the spin chain. In particular,
 the total momentum of an eigenstate is
 \begin{equation}\label{rapidities}
 P=\sum_{i=1}^M p_i\eqncom \hspace{0.5cm}\mbox{where}\hspace{0.5cm}  u_i=\frac{1}{2}\cot\frac{p_i}{2}\eqndot
 \end{equation}
 
As likewise explained in J.L.\ Jacobsen's lecture, the spin chain eigenstates can also be found by the coordinate-space Bethe
approach, which leads to the eigenstates being expressible as a sum over plane waves. More precisely, the 
highest-weight eigenstates with $M$ spins flipped compared to the reference state in this
approach take the following form
\begin{align}\label{eq:GenEigenstate}
|\vec{p}\rangle := |p_1,\ldots, p_M\rangle = \sum_{\sigma \in S_{M}} \sum_{1\leq n_1<\ldots<n_M\leq L} \e^{i
\sum_m (p_{\sigma_m} n_m + \frac{1}{2}\sum_{j<m} \theta_{\sigma_j\sigma_m} )} S^-_{n_1}\ldots S^-_{n_M} |0\rangle\eqncom
\end{align}
where the sum runs over all permutations and where the variables $p_1,\ldots,p_M$ now clearly have the interpretation of
the lattice momenta of the $M$ excitations (flipped spins). Moreover, up to an overall phase, the wave function \eqref{eq:GenEigenstate} only depends on the two-body S-matrix of the system 
\begin{align}
\mathcal{S}_{ij}  := \e^{\theta_{ij} - \theta_{ji}} = -\frac{1+ \e^{i p_i + i p_j} - 2 \e^{ip_i}}{1+ \e^{i p_i + i p_j} - 2 \e^{ip_j}}\eqndot
\end{align}
We are actually free to multiply \eqref{eq:GenEigenstate} with an arbitrary phase since this neither affects the spectrum nor the orthonormality of the Bethe vectors. We will fix the phase when we will discuss one-point functions in section \ref{onepoint}.

The momentum variables  are related to the 
rapidity variables as in~(\ref{rapidities})
but notice that the states $|\vec{p}\rangle$ and $|\textbf{u}\rangle$ are not identical but only proportional to each other.
The exact factor of proportionality was worked out in~\cite{Escobedo:2010xs}.  Whereas translational invariance tells us that the
total momentum constitutes a good quantum number, the cyclicity of the single-trace operators tells us that this quantum number
has to be a multiple of $2\pi$  for a Bethe state to qualify as a gauge-theory operator, i.e.\
\begin{equation}\label{cyclicity}
P=\sum_{k=1}^M p_k= 2\pi n\eqncom\hspace{0.5cm}\mbox{ or}\hspace{0.5cm} \prod_{k=1}^M\left(\frac{u_k+\frac{i}{2}}{u_k-\frac{i}{2}}\right)=1\eqndot
\end{equation}
In what follows, we will need the norm of a Bethe state \eqref{eq:GenEigenstate}. There is a elegant closed expression of determinant type due to Gaudin~\cite{Gaudin:1976sv}, see also~\cite{Korepin:1982gg}. 
Let us rewrite the Bethe equations \eqref{Betheeqns} and introduce the function $\Phi$ as their logarithm
\begin{align}
1 =\left(\frac{u_k-\frac{i}{2}}{u_k+\frac{i}{2}}\right)^L \prod_{\substack{j=1\\j\neq k}}^M \frac{u_k-u_j+i}{u_k-u_j-i} \equiv \exp[ i \Phi_k]\eqndot
\end{align}
Then, the norm is given in term of the Jacobian matrix $G_{ij} = \partial_{u_i} \Phi_j$:
\begin{align}\label{eq:Gaudin}
\langle \textbf{u} | \textbf{u} \rangle = \Big[ \prod_{i=1}^{M} u_i^2 + \frac{1}{4}\Big] \det G\eqndot
\end{align}
As mentioned above, the norm formula depends on the type of Bethe ansatz used and it will look different for the algebraic Bethe ansatz.

Now that we have described the eigenstates of the Heisenberg spin chain Hamiltonian, let us spell out the identification between spin-chain states and field-theory operators explicitly. Let $|\mathbf{u}\rangle$ be a Bethe state, then we can write 
\begin{equation}\label{eq:OviaU}
  \mathcal{O} \equiv  \left(\frac{1}{2g}\right)^{L}\frac{\mathcal{Z}}{\sqrt{L}}
    \frac{\tr\prod_{l=1}^L\Big(\langle\uparrow_l \!\! |\otimes X+\langle\downarrow_l \!\! |\otimes Y\Big) | \mathbf{u} \rangle }{\sqrt{ \langle \mathbf{u} | \mathbf{u} \rangle}}\eqncom
\end{equation}
where we have normalised the operators such that $M_{ij} = \delta_{ij}$. The explicit normalisation factor is most
easily derived from \eqref{eq:SU2-two-point-spin-chain} (since $|\mathbf{u}\rangle$ is an eigenstate, we effectively
have $H \to \Delta^{(1)}$). Requiring that the only one-loop correction to the two-point function is 
$\propto \log (x-y)^2$ (we work in units where $\mu = 1$) fixes the renormalisation constant to
\begin{equation}
  \renZ = 1 + g^2\frac{\Delta^{(1)}}{2}\left(\frac{1}{\peps} + 1 + \gammaE + \log\pi\right) + O(g^4) \eqndot
  \label{eq:Z-our-scheme}
\end{equation}

As an example, let us work out \eqref{eq:OviaU} for the Bethe states of length 2. The numerator of \eqref{eq:OviaU} becomes
\begin{align}
\langle\uparrow\uparrow\!|\mathbf{u}\rangle\, \tr [XX] + 
\langle\uparrow\downarrow\!|\mathbf{u}\rangle\, \tr [XY] +
\langle\downarrow\uparrow\!|\mathbf{u}\rangle\, \tr [YX] +
\langle\downarrow\downarrow\!|\mathbf{u}\rangle\, \tr [YY].
\end{align}
Then, for example for $|\textbf{u}\rangle =|0\rangle$ the corresponding operator becomes
\begin{align}\label{eq:OviaUexample}
  \mathcal{O}_{|0\rangle} \equiv  \frac{1}{4\sqrt{2}g^2}\, \tr [X^2]\eqncom
\end{align}
which has a unit-normalised two-point function as can be seen from \eqref{eq:SU2-two-point}.

Let us define a parity operation $\mathcal{P}$ which acts on single-trace operators by inverting the
orders of the fields inside the trace, i.e.\
\begin{equation}
\mathcal{P} \cdot\tr[\phi_{i_1}\phi_{i_2}\cdots\phi_{i_L}] =\tr[\phi_{i_L}\phi_{i_{L-1}}\cdots\phi_{i_1}]\eqndot
\end{equation}
Obviously, the Heisenberg Hamiltonian commutes with the parity operation.  This means that the
spin-chain eigenstates can be chosen to be parity eigenstates as well. However, the spin-chain eigenstates generated by the
algebraic Bethe ansatz are not parity eigenstates as parity anti-commutes with all the odd charges, in particular $Q_3$, i.e.
\begin{equation}
[H,\mathcal{P}]=0\eqncom \hspace{0.5cm} \{ Q_3,\mathcal{P}\}=0\eqndot
\end{equation}
As usual, the parity operation changes the sign of all momenta and hence the sign of of all rapidities and it squares to the
identity. Let us denote by $|-\textbf{u}\rangle$ the state given by the right-hand side of (\ref{eigenstate}) with each $u_i$ being replaced by $-u_i$.
Then, $\mathcal{P}|\textbf{u}\rangle$ can differ from $|-\textbf{u}\rangle$ by at most a phase factor.
Since the Bethe equations~(\ref{Betheeqns}), the cyclicity constraint~(\ref{cyclicity}) 
and the expression for the energy~(\ref{Energy}) are all invariant under $u_i \rightarrow -u_i$,
the state $|-\textbf{u}\rangle $ is again a cyclically invariant eigenstate of $\mathcal{H}$ with the same eigenvalue 
as $|\textbf{u}\rangle $.
It thus follows that the eigenstates of $\mathcal{H}$
can be separated into unpaired states for which $|\textbf{u}\rangle =|-\textbf{u}\rangle$ and paired states
$(|\textbf{u}\rangle, |-\textbf{u}\rangle)$ for which $|\textbf{u}\rangle \neq |-\textbf{u}\rangle$. The unpaired states will play a distinguished
role in section~\ref{onepoint}. These states are parity eigenstates and can be shown to fulfil
\begin{equation}
\mathcal{P} |\textbf{u}\rangle_{\text{unpaired}}= (-1)^{M(L+1)} |\textbf{u}\rangle_{\text{unpaired}}\eqncom \hspace{0.5cm}
 Q_3 |\textbf{u}\rangle_{\text{unpaired}}=0\eqndot 
\end{equation}
The degenerate states in a parity pair are not states of a definite parity but can be combined into parity eigenstates,
so-called parity pairs $(|+\rangle,|-\rangle)$, in the following way:
\begin{equation}
|+\rangle= |\textbf{u}\rangle+|-\textbf{u}\rangle\eqncom \hspace{0.5cm} |-\rangle= |\textbf{u}\rangle-|-\textbf{u}\rangle\eqncom
\end{equation}
where
\begin{equation}
\mathcal{P} |+\rangle= (-1)^{M(L+1)} |+\rangle\eqncom \hspace{0.5cm}
\mathcal{P} |-\rangle= -(-1)^{M(L+1)} |-\rangle\eqncom
\end{equation}
and
\begin{equation}\label{eq:QP}
Q_3 |+\rangle\propto|-\rangle\eqncom \hspace{0.5cm} Q_3 |-\rangle\propto|+\rangle\eqndot
\end{equation}

\subsection{Higher loop orders}
\begin{figure}
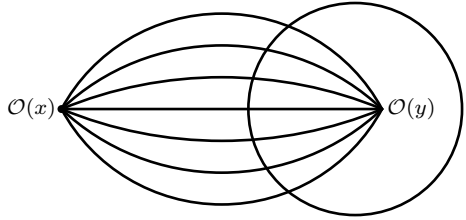

\centering
  \fmfframe(20,20)(20,20){%
 \begin{fmfchar*}(120,40)
 \fmfleft{o1}
 \fmfright{i1}
 \fmf{plain,left=0.6}{i1,o1}
 \fmf{plain,left=0.4}{i1,o1}
 \fmf{plain,left=0.2}{i1,o1}
 \fmf{plain,left=0.0}{i1,o1}
 \fmf{plain,left=-0.2}{i1,o1}
 \fmf{plain,left=-0.4}{i1,o1}
 \fmf{plain,left=-0.6}{i1,o1}
 \fmffreeze
 \fmfposition
 \fmfiv{label=$\scriptstyle \mathcal{O}(x)$,l.dist=2,decor.shape=circle,decor.filled=full,
decor.size=thick}{vloc(__o1)}
 \fmfiv{label=$\scriptstyle \mathcal{O}(y)$,l.dist=2,decor.shape=circle,decor.filled=full,
decor.size=thick}{vloc(__i1)}
 \fmfiv{decor.shape=circle,decor.filled=empty,
decor.size=40thick}{vloc(__i1)-(10,0)}
 \end{fmfchar*}
 }
\caption{Example of an interaction wrapping once around the operator.}
\label{wrapping}
\end{figure}

By doing  explicit higher-loop computations following the same strategy as in section~\ref{spectrum}, one can likewise derive a perturbative expression for the
dilatation operator, i.e.\
\begin{align}
  D = \sum_{n=0}^\infty g^{2n} D^{(n)}\eqndot
  \label{eq:Dilop-expansion}
\end{align}
For the two-loop contribution, one finds~\cite{Beisert:2003tq}
\begin{equation}\label{twoloop}
D^{(2)}=
-2\sum_{i=1}^L (\mathbb{P}_{i,i+2}-1) + 8\sum_{i=1}^L(\mathbb{P}_{i,i+1}-1)
\eqndot
\end{equation}
We notice, in particular, that the anomalous dimension of the BPS operator $\tr X^L$ stays zero at two-loop order as
expected.
When one  diagonalised  the dilatation operator including this correction term initially (by brute force), one observed that to order $g^2$ the spectrum still contained the same number of pairs of degenerate eigenstates with opposite parity. This fact 
was viewed as a smoking gun of higher-loop integrability as it hinted at the continued existence of a conserved third 
charge~\cite{Beisert:2003tq}.
Indeed, it is possible to perturbatively modify the third and the higher order charges by terms of order $g^2$, i.e.
\begin{equation}
Q_i= Q_i^{(0)}+g^2 Q_i^{(1)}\eqncom
\end{equation} 
in such a way that the quantum-corrected charges commute up to terms of order $g^4$:
\begin{equation}
[Q_i,Q_j]= O(g^4)\eqndot
\end{equation}
The same idea can be pursued at general loop order $\ell$ where one would have
\begin{equation}
Q_i=Q_i^{(0)}+g^2 Q_i^{(1)}+g^4 Q_i^{(2)}+\ldots+ g^{2\ell}Q_i^{(\ell)}\eqncom \hspace{0.5cm} [Q_i,Q_j]= O(g^{(2\ell+2)})\eqncom
\end{equation}
and one would denote the system as being perturbatively integrable.
Here, the correction $Q_i^{(\ell)}$ is an operator which involves $i+\ell$ neighbouring spins. Concretely, the idea has been implemented
up to four-loop order~\cite{Beisert:2007hz} . The algebraic Bethe ansatz approach does not apply to the quantum corrected system, where
the interaction is no longer of nearest-neighbour type. Nevertheless, a modified, so-called asymptotic Bethe ansatz exists. It has been argued for in a long series a papers where the focus was shifted from the Hamiltonian and the conserved charges to the
two-body scattering matrix of the theory~\cite{Staudacher:2004tk} 
and the calculational effort was shifted from brute-force field-theoretical computations to symmetry considerations.  The asymptotic Bethe equations read~\cite{Beisert:2005fw}
\begin{align}\label{AsBetheeqns}
\left(\frac{x(u_k+\frac{i}{2})}{x(u_k-\frac{i}{2})}\right)^L =\prod_{\substack{j=1\\j\neq k}}^M \frac{u_k-u_j+i}{u_k-u_j-i}\,
\exp(2i\theta(u_k,u_j))\eqncom
 \end{align}
 where $\theta(u_k,u_j)$ is denoted as the dressing phase and is explicitly known~\cite{Vieira:2010kb}. The Zhukovski 
 variable $x(u)$ is defined via
 \begin{equation}\label{xu}
 u=x+\frac{g^2}{x}\eqndot 
 \end{equation}
 Furthermore, the cyclicity condition now reads
 \begin{equation}
\prod_{k=1}^M\left(\frac{x(u_k+\frac{i}{2})}{x(u_k-\frac{i}{2})}\right)=1\eqncom
 \end{equation}
 and the expression for the energy eigenvalues is modified to
 \begin{equation}\label{AsEnergy}
E=\sum_{j=1}^M i\left(\frac{1}{x(u_j+\frac{i}{2})}-\frac{1}{x(u_j-\frac{i}{2})}\right)\eqndot
 \end{equation}
 The dressing phase only plays a role at four-loop order and beyond.  The Bethe equations~(\ref{AsBetheeqns}) are asymptotic
 in the sense that they are only valid when expanded perturbatively to a given order $n$ in $g^2$ for operators whose length
 is smaller than or equal to $n$. If this criterion is not fulfilled, one has to take into account wrapping corrections which, as indicated by their name, are corrections which occur when the spin-chain interaction wraps once or more around the 
 operator~\cite{Ambjorn:2005wa}, see figure~\ref{wrapping}.

The asymptotic Bethe equations give access to the spectrum at higher loops but not to the corresponding wave functions.
Wave functions at higher loop orders can be generated by a technique known as $\Theta$-morphism~\cite{Gromov:2012uv}.

\subsection{Beyond the planar limit}

The derivation in section~\ref{spectrum} can be generalised to give the full one-loop dilatation operator and not only its large-$N$ limit. Including
all terms, the action of the one-loop dilatation operator in the $SU(2)$ sector can be expressed in terms of an effective vertex, 
acting on an operator \cite{Beisert:2002bb}:
\begin{equation}
V=\sum_{n=0}^{\infty} g^{2n} V^{(n)}\eqndot
\end{equation}
The one-loop contribtion reads
\begin{equation} \label{effective_vertex}
V^{(1)}= -\frac{2}{N} :\mbox{tr} [X,Y][\check{X},\check{Y}]:\eqncom \hspace{0.5cm} (\check{X})_{ab}=\frac{\delta}{\delta X_{ab}}\eqncom
\end{equation}
where the normal ordering symbol signifies that the derivatives are not allowed to act on fields belonging to the effective vertex itself. Going beyond the large-$N$ limit, one cannot restrict one-self to single-trace operators but has to consider also multi-trace
ones since the action of the dilatation operator now leads to splitting and joining of traces as illustrated by the example
below.
Notice that we only show one out of four terms contributing
to the dilatation generator and only one possible way of 
applying the derivatives:
 
\vspace{-\baselineskip} 
\ifarxiv 
\setlength{\unitlength}{1.18\unitlength}
\fi 
\begin{picture}(420,50)(0,0)
\put(0,0){$\tr(XY \check{X}\check{Y}) \cdot
\tr(Y X Y Y X )\, \tr(Y X)=
\tr(X Y \check{X} X YY X)\, \tr(YX)$}
\put(42,20){\line(1,0){30}}
\put(42,20){\line(0,-1){7}}
\put(72,20){\line(0,-1){7}}
\put(197,17){\line(1,0){6}}
\put(197,17){\line(0,-1){4}}
\put(203,17){\line(0,-1){4}}
\put(206,13){\tiny 1}
\put(195,19){\line(1,0){35}}
\put(195,19){\line(0,-1){6}}
\put(230,19){\line(0,-1){6}}
\put(233,13){\tiny 2}
\put(193,21){\line(1,0){71}}
\put(193,21){\line(0,-1){8}}
\put(264,21){\line(0,-1){8}}
\put(270,13){\tiny 3}
\put(10,-30){$=N\tr(XYYYX)\,\tr(YX)+ 
\tr(XY)\, \tr(XYY)\,\tr(YX)+\tr(XYXXXYYX)\eqndot$}
\end{picture}
\vspace*{1.2cm}

From this example, it should be clear that we can decompose the vertex representing the
full one-loop dilatation operator  for finite $N$ in the following way:
\begin{equation}
V^{(1)}= D^{(1)}+\frac{1}{N} \,D^{(1)}_{+}+\frac{1}{N} \,D^{(1)}_{-}\eqncom
\end{equation}
where $D^{(1)}$, which was given in~(\ref{eq:HfromD}), conserves the number of traces, $D^{(1)}_{+}$ increases the trace number by one and $D^{(1)}_{-}$
reduces the trace number by one. In the language of spin chains, $D^{(1)}_{+}$ splits a chain into two parts while 
$D^{(1)}_{-}$ joins two chains into one. 
 In a similar manner, the full non-planar two-loop contribution to the dilatation operator can be expressed in terms of an effective vertex
as~\cite{Beisert:2003tq} 
\begin{equation}
V^{(2)}= D^{(2)}+\frac{1}{N} \,D^{(2)}_{+}+\frac{1}{N} \,D^{(2)}_{-}+\frac{1}{N^2} \,D^{(2)}_{++}+\frac{1}{N^2} \,D^{(2)}_{--}\eqncom
\end{equation}
where $D^{(2)}$ was given in~\eqref{twoloop} and
where $D^{(2)}_{++}$ increases the trace number by two and $D^{(2)}_{--}$ reduces the trace number by two. 
\begin{exc}
By applying the effective vertex~(\ref{effective_vertex}) to an operator from the $SU(2)$ sector, show that it 
reduces to the Hamiltonian of the Heisenberg spin chain in the limit $N\rightarrow \infty$.
\end{exc}
The splitting and joining of traces or spin chains constitute highly non-local interactions for which the traditional tools of integrability are not applicable. One naive thing that one can do is to consider a finite, closed set of multi-trace operators
of a given length $L$ and with a given number of excitations, $M$, and diagonalise the dilatation operator including 
its non-planar terms by brute force in this subspace. Alternatively, one can at a slightly more advanced level start by
diagonalising the planar part of the Hamiltonian, still in a finite, closed set of multi-trace operators, treat the $\frac{1}{N}$ terms in the dilatation operator as a perturbation and do quantum-mechanical perturbation theory in $\frac{1}{N}$.
From these types of simple analyses, there are few things that one can learn~\cite{Beisert:2003tq}.
 First, one observes that the degeneracy between
the planar parity pairs (the ($|+\rangle$, $|-\rangle$) states) gets lifted when $\frac{1}{N}$ corrections are taken into account.
Thus, the smoking gun of integrability is no longer present.  Furthermore, for states which are degenerate at the planar 
level, such as the planar parity pairs,
the leading non-planar correction to the energy behaves as $\frac{1}{N}$, whereas in the generic case the first non-planar
correction behaves as $\frac{1}{N^2}$. This is a simple consequence of quantum-mechanical perturbation theory. Finally,
starting at the two-loop level, one finds that there are states which do not have a well-defined double expansion in 
$g^2$ and $\frac{1}{N}$.

 Aiming at going beyond the planar level in a more systematic approach, a convenient basis of $SU(2)$ operators might be the so-called restricted Schur polynomials, which  constitute a basis of  multi-trace operators that are orthogonal for finite $N$. Studying the action of the one- and two-loop dilatation operator in this basis of operators, it is possible by imposing various
limits on top of the large-$N$ limit to find an integrable sub-system which, however, looks like a set of decoupled harmonic oscillators~\cite{Carlson:2011hy}.

Another direction of investigation, which can be viewed as the first step in going beyond the planar level, is the study of
three-point functions as these can be seen as building blocks for non-planar correlation functions. Whereas in the study of the spectral problem of ${\mathcal N}=4$ SYM theory one only
needs the eigenvalues of the spin-chain Hamiltonian, in the study of the theory's three-point functions the explicit form of the  spin-chain eigenfunctions plays a crucial role.
The study of three-point
functions has recently been boosted by the development of the so-called hexagon-program, which is covered in the lectures by
S.\ Komatsu.

\section{\texorpdfstring{$\mathcal{N}=4$}{N=4} SYM theory with a defect, one-point functions and integrability}
\label{onepoint}

Rather than computing three-point functions in $\mathcal{N}=4$ SYM theory, we will focus on a related problem which also requires the knowledge of the explicit form of the Bethe 
wave functions.
We will compute the one-point functions in a certain defect version of $\mathcal{N}=4$ SYM theory.

\subsection{\texorpdfstring{$\mathcal{N}=4$}{N=4} SYM theory with a defect}
\label{sec: introtoDCFT}

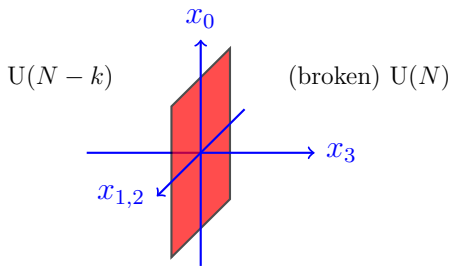
\begin{figure}[h]
\begin{center}
\scalebox{1}{
\begin{tikzpicture}
	[	axis/.style={->,blue,thick},
		axisline/.style={blue,thick},
		cube/.style={opacity=.7, thick,fill=red}]

	\draw[axisline] (-1.5,0,0) -- (0,0,0) node[anchor=west]{};	
		
	\draw[cube] (0,-1,-1) -- (0,1,-1) -- (0,1,1) -- (0,-1,1) -- cycle;

	\draw[axis] (0,0,0) -- (1.5,0,0) node[anchor=west]{$x_3$};
	\draw[axis] (0,-1.5,0) -- (0,1.5,0) node[anchor=south]{$x_0$};
	\draw[axis] (0,0,-1.5) -- (0,0,1.5) node[anchor=east]{$x_{1,2}$};	
		
	\node[anchor=east] at (-1,1,0) {\scalebox{0.8}{$\mathrm{U}(N-k)$}};
	\node[anchor=west] at (1,1,0) {\scalebox{0.8}{(broken) $\mathrm{U}(N)$}};
	
\end{tikzpicture}
}
\caption{The defect theory.}\label{fig:dCFT}
\end{center}
\end{figure}

There exists a certain defect version of $\mathcal{N}=4$ SYM theory for which half of the supersymmetries are preserved and for which a holographic dual exists. In this theory,
a codimension-one defect is positioned at $x_3=0$ and divides space into two regions, $x_3>0$ and $x_3<0$. In the bulk, one still has $\mathcal{N}=4$ SYM theory but with different
gauge groups on the two sides of the defect. The gauge group for $x_3<0$ is  $U(N-k)$, while the gauge group for $x_3>0$ is $U(N)$, see figure \ref{fig:dCFT}. The $U(N)$ 
symmetry for $x_3>0$, however, is broken by some of the scalar fields acquiring a non-trivial vacuum expectation value (vev) so that the gauge symmetry there is also effectively
$U(N-k)$. Due to the non-vanishing vevs, one-point functions can be non-trivial on one side of the defect already at tree level.
In addition to the usual action of $\mathcal{N}=4 $ SYM theory, the system has a three-dimensional action involving fields that are confined to the defect. These defect fields have self-interactions as well
as interactions with the bulk fields of $\mathcal{N}=4 $ SYM theory \cite{DeWolfe:2001pq,Erdmenger:2002ex}. In the remainder of the lectures, we will only work at tree level and at one-loop order where the defect field theory does not come into play.

\subsubsection{Symmetries}
\label{sec: symmetry with defect}
Introducing the codimension-one defect at $x_3=0$ breaks several of the original symmetries of $\mathcal{N}=4$ SYM theory discussed in section \ref{sec: symmetries}.
To start with, let us analyse the minimal possible consequences of introducing the defect.
The condition $x_3=0$ is preserved by the translations $P_{\hat\mu}$, $\hat\mu=0,1,2$, but not by $P_3$.
Similarly, the Lorentz transformations $M_{\hat\mu\hat\nu}$ preserve $x_3=0$, but $M_{\hat\mu3}=-M_{3\hat\mu}$ does not. The four-dimensional Poincar\'{e} symmetry is thus reduced to three-dimension Poincar\'{e} symmetry.
A scale transformation $D$ preserves $x_3=0$ and so do the special conformal transformations $K_{\hat\mu}$ but not $K_3$. The four-dimensional conformal group $SO(4,2)\simeq SU(2,2)$ is thus reduced to the three-dimensional conformal group $SO(3,2)\simeq Sp(4)$.

While this analysis is straightforward in vector indices, let us now redo it in spinor indices as a preparation for understanding the influence on supersymmetry. For convenience, we will assume here that the defect is at $x_2=0$ instead of $x_3=0$.
\begin{exc}
 Work out the similarity transformation that relates these cases.
\end{exc}
\noindent
Via the Pauli matrices $\sigma^\mu_{\alpha\dot\alpha}$, the Lorentz vector $P_\mu$ is translated to a $2\times 2$ matrix. The condition that the component $P_2$ vanishes then translates to the matrix being symmetric.  
We thus have to determine the Lorentz transformations that yield symmetric matrices when applied to symmetric matrices.
They are given by $\hat{L}^\alpha{}_\beta={L}^\alpha{}_\beta+\dot{L}^{\dot\alpha}{}_{\dot\beta}$.\footnote{Here, we identify dotted and undotted indices, so for example $\hat{L}^1{}_2={L}^1{}_2+\dot{L}^{\dot1}{}_{\dot2}$.}
This explicitly shows how the four-dimensional Lorentz group $SO(1,3)\simeq SU(2)_L\times SU(2)_R$ is reduced to the three-dimensional Lorentz group $SO(1,2)\simeq SU(2)$.

Recalling that supercharges anticommute to translations \eqref{eq: QQ P}, some of the supersymmetry is necessarily broken as well.
As the supercharges have spinor indices, we now benefit from the previous analysis in spinor indices.
First, we observe that the preserved supercharges have to be spinors of $\hat{L}$.
Second, they have to anticommute to a symmetric matrix in the spinor indices.
This leads to half of the supercharges being preserved, namely $\hat{Q}_{\alpha}^A={Q}_{\alpha}^A+\dot{Q}_{\dot\alpha}^A$.  From this choice, we see that the anti-commutator of the $\hat{Q}$s is proportional to a symmetrised version of the momentum $P$, which does not contain $P_2$.
In the same way, only half of the superconformal charges preserve $x_2=0$, as they anticommute to special conformal transformations.
The preserved supercharges are manifestly real. Thus, the R-symmetry group $SU(4)\simeq SO(6)$ that acts on them is reduced to  $SO(4)\simeq SO(3)\times SO(3)$.
In total, the superconformal group $PSU(2,2|4)$ of $\mathcal{N}=4$ SYM theory is thus reduced to $OSP(4|4)$.

So far, we have only considered the minimal effect of introducing a codimension-one defect into $\mathcal{N}=4$ SYM theory. Depending on which fields occur on the defect and how they interact among themselves and with the fields of $\mathcal{N}=4$ SYM theory, also more symmetry could be broken.
However, there does indeed exist a defect action such that the (quantum) theory preserves $OSP(4|4)$, at least for $k=0$ \cite{DeWolfe:2001pq}.

\subsubsection{Correlation functions}
The correlation functions in a CFT with a boundary or a codimension-one defect are less restricted than for a usual CFT. This is due to the fact that a defect breaks part of the conformal symmetry, as just discussed. 
Already one-point functions of composite operators $\cO_i$ can be non-vanishing. The remaining conformal symmetry and the scaling dimension $\Delta_i$ of the operator fix the one-point functions up to a constant $a_i$ \cite{Cardy:1984bb}:
\begin{equation}
\label{eq: form of the one-point function}
 \langle \mathcal{O}_i(x)\rangle=\frac{a_i}{x_3^{\Delta_{i}}}\eqndot
\end{equation}
We see that one-point functions in a dCFT exhibit a complexity similar to three-point functions in a CFT.
Two-point functions in a dCFT can be non-vanishing also for operators of unequal scaling dimensions and are fixed to be of the form
\begin{align}
\langle \mathcal{O}_i(x) \mathcal{O}_j(y) \rangle = \frac{f(\xi)}{x_3^{\Delta_i}y_3^{\Delta_j}}\eqncom
\end{align}
where $f(\xi)$ is a function of the conformal ratio $\xi=\frac{|x -y|^2}{4 x_3 y_3}$.

Finally, all correlation functions in a dCFT should reduce to the corresponding correlation functions in the absence of the defect if the distance to the defect is large compared to the distance between the insertion points.

\subsubsection{Vacuum expectation values}

For our specific model,  the vacuum expectation values that the scalar fields pick up  are described by $SU(2)$ representations \cite{Constable:1999ac}. More precisely,
for $x_3>0$
\begin{align}\label{eq:vev}
&\phi _i^{\rm cl} = -
 \frac{1}{x_3}\,\begin{pmatrix}
  \left(t_i\right)_{k\times k} & 0_{k\times (N-k)}\\
 \hspace*{0.5cm} 0_{(N-k)\times k} & \hspace*{0.7cm}0_{(N-k)\times (N-k)} \\
 \end{pmatrix}\eqncom&&i=1,2,3\eqncom\\
&\phi ^{\rm cl}_i=0\eqncom&& i=4,5,6\eqncom
\end{align}
where the three $k\times k$ matrices $t_i$, $i=1,2,3$  constitute a $k$-dimensional unitary, irreducible representation of $SU(2)$;
in particular,
\begin{equation}\label{eq:su2relations}
 \left[t_i,t_j\right]=i\varepsilon _{ijk}t_k\eqndot
\end{equation}
For $x_3<0$, all classical fields are vanishing.

\subsubsection{Representation of the algebra of \texorpdfstring{$SU(2)$}{SU(2)}} To be explicit, let us here spell out the $k$-dimensional irreducible representation. Introduce the standard $k\times k$ matrix unities $E^i{}_j$ that are zero everywhere except for a 1 at position $(i,j)$. These matrices satisfy the relation $E^i{}_{j}E^k{}_{l} = \delta^k{}_j E^i{}_l$.

Next, we consider the following constants
\begin{align}\label{eq:su2rep}
&c_{i} = \sqrt{i(k-i)}\eqncom
&& d_{i} = \frac{1}{2}(k-2i+1)\eqncom
\end{align}
together with the following matrices
\begin{align}\label{eq:tgenerators}
& t_+= \sum_{i=1}^{k-1} c_{i} E^i{}_{i+1}\eqncom
&& t_- =\sum_{i=1}^{k-1}  c_{i} E^{i+1}{}_{i}\eqndot
\end{align}
The usual $k$-dimensional $SU(2)$ representation is then given by
\begin{align}\label{eq:defrep}
&t_1 = \frac{t_+ +t_-}{2}\eqncom
&&t_2 = \frac{t_+ -t_-}{2i}\eqncom
&& t_3 =\sum_{i=1}^k  d_{i} E^{i}{}_{i}\eqndot
\end{align}
It is easy to check that these matrices satisfy the commutation relations \eqref{eq:su2relations}. 
For the important special case $k=2$, the representation matrices are multiples of the Pauli matrices: $t_i|_{k=2}=\frac{1}{2}\sigma_i$.

\subsection{Tree-level one-point functions in the \texorpdfstring{$SU(2)$}{SU(2)} sector \label{tree-level-onepoint}}

\begin{figure}[t]
 \centering
  \includegraphics{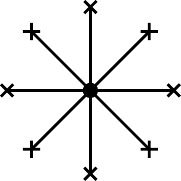}
  \caption{The tree-level one-point function is obtained by inserting the classical solution into the operator. The operator is depicted by a dot and the insertion of the classical solution by a line with a cross at the end.}  \label{fig: tree-level one-point}
\end{figure}

At tree level, the one-point function is obtained by inserting the classical solution \eqref{eq:vev} in an operator, see figure \ref{fig: tree-level one-point}. Clearly, only operators consisting solely of scalar operators can have a non-zero one-point function. In what follows, we will work in the planar limit so that we can apply the integrability techniques that were previously discussed. This means that we restrict to single-trace operators of the form
\begin{align}
\mathcal{O} = \Psi^{i_1\ldots i_L} \tr( \scal_{i_1} \ldots \scal_{i_L})\eqndot
\end{align}
Inserting \eqref{eq:vev} into such an operator $\cO$ then gives us at tree level
\begin{align}\label{eq:vevO}
\langle \mathcal{O} \rangle^{cl} = (-1)^L\Psi^{i_1\ldots i_L}\frac{ \tr( t_{i_1} \ldots t_{i_L})}{x_3^L}\eqndot
\end{align}
For any given operator, the above expression can straightforwardly be evaluated. However, this is hardly a constructive approach. For instance, to compute the one-point function of a scalar operator corresponding to a Bethe state, we would have to write out its explicit wave function. 
Instead, we will now derive by integrability techniques 
 a closed formula for $\langle \mathcal{O} \rangle^{cl}$ which is expressed entirely in terms of  the length, the number of excitations and
 the specific Bethe roots characterising the operator.

\subsubsection{The matrix product state} From now on, let us restrict to the $SU(2)$ sector. The first step to a more systematic approach is the realisation that \eqref{eq:vevO} can be written as an inner product between the Bethe state $|\mathbf{u}\rangle$ corresponding to our operator via \eqref{eq:OviaU} and a so-called matrix product state (MPS):
\begin{align}
 |\MPS\rangle_L= \tr \prod_{n=1}^L  \Big[t_1 \otimes |\!\uparrow\rangle_n  +  t_2 \otimes|\!\downarrow\rangle_n\Big]\eqndot
\end{align}
The subscript $n$ stands for the usual embedding in the $L$-fold tensor product, while the trace is as usual in colour space.  The MPS depends on the length $L$ of the spin chain that we are considering, but in order to avoid cumbersome notation, we will from now on omit the subscript $L$.

\begin{exc} 
Compute the MPS for $L=2,4$.
\end{exc}

\noindent
 
Using the explicit relation between the Bethe states and the field-theory operators  \eqref{eq:OviaU},
the problem of computing a one-point function then reduces to computing the following quantity
\begin{align}
&\langle\mathcal{O}\rangle^{cl} = (-1)^L\left(\frac{1}{2g}\right)^{L}\frac{\mathcal{Z}}{\sqrt{L}} \frac{C_k}{x_3^L}\eqncom
&&C_k = \frac{\langle\MPS|\mathbf{u} \rangle}{\sqrt{\langle\mathbf{u} |\mathbf{u} \rangle}}\eqncom
\end{align}
where the various proportionality factors ensure that the operator is properly normalised. 

There is the important subtlety that $C_k$ is only defined up to a phase. In our identification of the field-theory operator and Bethe state, we can always insert an additional phase factor. This obviously leaves the two-point function invariant, but it will affect the overlap with the MPS. In order to fix this ambiguity, we will always choose the overall phase such that $C_k$ is real and positive.

\subsubsection{Generalities}

It is easy to show that $C_k$ is only non-vanishing if both $L$ and $M$ are even, where $M$ is the number of 
excitations or equivalently the number of Bethe roots. Namely, the Lie algebra of $SU(2)$ admits an isomorphism where two of the $t$'s are mapped to $-t$. This isomorphism is realised by a similarity transformation which leaves the MPS invariant due to cyclicity of the trace. For example, consider the case when $(t_1,t_2,t_3)\rightarrow(-t_1,-t_2,t_3)$. This immediately implies that 
\begin{align}
\langle\MPS|\mathbf{u} \rangle = (-1)^L\langle\MPS|\mathbf{u} \rangle\eqncom
\end{align}
which means that $L$ has to be even. Similarly, it follows that $M$ has to be even. For details, we refer to~\cite{deLeeuw:2015hxa}.

Apart from these restrictions on the quantum numbers, for a non-zero overlap with the MPS we also need some restrictions on the Bethe roots. It is easy to see that the MPS is parity even, so only parity even Bethe states can have a non-trivial overlap with the MPS. Finally, it can be shown that $Q_3$ annihilates the MPS~\cite{deLeeuw:2015hxa}. From \eqref{eq:QP}, we then see that the only possible states that have a non-vanishing one-point functions are states that satisfy $|\mathbf{u}\rangle = |-\mathbf{u}\rangle$.

\subsubsection{Vacuum} The first state to consider is the ferromagnetic vacuum \eqref{eq: vacuum}, which corresponds to the operator $\tr X^L$.
Its one-point function is given by
\begin{align}
C_k = \frac{\langle\MPS|0\rangle}{\sqrt{\langle0|0\rangle}} = \tr( t_1^L) = \sum_{i=1}^k d_i^L = -2\frac{B_{L+1}(\frac{1-k}{2})}{L+1}\eqncom
\end{align}
where $d_i$ are the coefficients defining the $SU(2)$ representation \eqref{eq:su2rep} and $B_{L+1}$ is the Bernoulli polynomial with index $L+1$. We see that the one-point function is a polynomial in $k$ of degree $L+1$.

\subsubsection{One-point functions for \texorpdfstring{$k=2$}{k=2}}

The simplest case that we can consider for $M>0$ is the case $k=2$. This actually turns out to be a fundamental building block for the general $k$ case. For $k=2$, the $t$-matrices are simple multiples of  the Pauli matrices: $t_i=\frac{1}{2}\sigma_i$. They satisfy the following relations:
\begin{align}
&t_i^2 = \frac{1}{4}\eqncom
&& t_i t_j = -t_j t_i  \quad\mathrm{for}\quad i\neq j.
\end{align}
This means that the inner product of the MPS with a Bethe state \eqref{eq:GenEigenstate} dramatically simplifies. In particular, any trace factor can be easily evaluated: 
\begin{align}
\tr ( t_1^{n_1-1}t_2 t_1^{n_2-n_1-1} t_2\ldots ) = (-1)^{n_1+n_2+\ldots} \tr( t_1^{L-M}t_2^{M}) = 2^{1-L}(-1)^{n_1+n_2+\ldots} \eqndot
\end{align}
Thus, the inner product of a Bethe state with the MPS takes the following form:
\begin{align}
\langle\MPS|\mathbf{u}\rangle = 2^{1-L}\sum_{\sigma \in S_{M}} \sum_{n_i} \e^{i
\sum_m (p_{\sigma_m} n_m + \frac{1}{2}\sum_{j<m} \theta_{\sigma_j\sigma_m} )} (-1)^{n_1 + \ldots +n_M}\eqndot
\end{align}
\begin{exc}
For two particles, compute the overlap $\langle\MPS|u,-u\rangle$ and show that upon using the Bethe equations \eqref{Betheeqns} it becomes
\begin{align}
\langle\MPS|u,-u\rangle = \frac{L}{2^{L-1}} \sqrt{\frac{u^2 + \frac{1}{4}}{u^2}}\eqndot
\end{align}
Also show that $\langle u,-u|u,-u\rangle = L(L-1)$ such that $C_2 = \frac{1}{2^{L-1}} \sqrt{\frac{u^2 + \frac{1}{4}}{u^2}}\sqrt{\frac{L}{L-1}}$.
\end{exc}

\noindent To describe the overlap for a general number of excitations $M$, we introduce the following function
\begin{align}
K_{ij} : = \frac{1}{2} \left[ \frac{1 + 4 u_i^2} {1 + (u_i + u_j)^2} + \frac{1 + 4 u_i^2} {1 + (u_i - u_j)^2}  \right]\eqncom
\end{align}
and the following $\frac{M}{2}\times \frac{M}{2}$ matrix
\begin{align}
F_{ij} : = \left(L - \sum_{n=1}^{M/2} K_{in}\right) \delta_{ij} + K_{ij}\eqndot
\end{align}
The overlap is then given by
\begin{align}\label{eq:overlap}
\langle\MPS|\mathbf{u}\rangle_{k=2} = 2^{1-L}
(\det F) \sqrt{\prod_{i=1}^{M/2} \frac{u^2_i +\frac{1}{4}}{u^2_i}}\eqndot
\end{align}
In order to finally obtain the one-point function $C_2$, we need to divide by the norm of the Bethe state \eqref{eq:Gaudin}. For states with paired rapidities $|\textbf{u}\rangle = |-\textbf{u}\rangle $, the norm formula factorises. Let us order the roots as $\{u_1,\ldots, u_{\frac{M}{2}}, -u_1,\ldots, -u_{\frac{M}{2}}\}$ and introduce the following $\frac{M}{2}\times\frac{M}{2}$ dimensional matrices $G_{\pm}$:
\begin{align}
\label{eq: definition of G pm}
&G_{\pm} = \partial_{u_m} \Phi_n \pm \partial_{u_{m+\frac{M}{2}}} \Phi_n\eqncom
\end{align}
then $\det G = \det G_+ \det G_-$. In terms of these matrices, the one-point function for $k=2$ can finally be written as
\begin{align}\label{eq:SU2quotient}
C_2 = 2^{1-L}\sqrt{ \frac{Q(\frac{i}{2})}{Q(0)}}\sqrt{ \frac{\det  G_+}{\det  G_-}} \eqncom
\end{align}
where  $Q(u) = \prod_{i=1}^M (u-u_i)$ is the Baxter polynomial. This means in particular that $\langle\MPS|\mathbf{u}\rangle_{k=2} \sim \det G_+$. It is an interesting open question whether there is a state $|A\rangle$ such that $\langle A|\mathbf{u}\rangle_{k=2} \sim \det G_-$.

A formula of the same type as  (\ref{eq:overlap}) has been obtained for the eigenstates of the $SU(3)$ spin chain  \cite{deLeeuw:2016umh}, and this result  has recently found application in the study of quantum quenches 
\cite{Calabrese:2017xx}. Furthermore, the expression (\ref{eq:SU2quotient}) can be generalised to 
tree-level one-point functions of the $SU(3)$ sector of 
${\cal N}=4$ SYM theory \cite{deLeeuw:2016umh}. We note, however, that the $SU(3)$ sector is a closed sector only to one-loop order.

\subsubsection{N\'eel state}

There is actually an interesting relation of one-point functions $C_2$ to the condensed-matter literature. It turns out that the MPS is cohomologically equivalent to the so-called N\'eel state:
\begin{align}
|\Neel\rangle = |\! \uparrow\downarrow\uparrow\downarrow\ldots\rangle + |\! \downarrow\uparrow\downarrow\uparrow\ldots\rangle\eqndot
\end{align}
The N\'eel state is a state at half-filling, i.e.\ it has $M=L/2$. It can be shown \cite{deLeeuw:2015hxa} that 
\begin{align}
2^L \Big(\frac{i}{2}\Big)^M|\MPS \rangle \Big|_{M=L/2}= |\Neel\rangle  + S^- |\ldots\rangle\eqndot
\end{align}
One of the remarkable properties of the Bethe ansatz is that the Bethe states are highest-weight states. This means that $S^+ |\mathbf{u}\rangle =0$ and thus for any Bethe state with $M=L/2$ the overlap of the MPS is the same as the overlap of the Bethe state with the N\'eel state, i.e.\ $2^{L-M} i^M\langle \mathbf{u}|\MPS \rangle= \langle \mathbf{u}|\Neel\rangle$. This is a problem that has been studied in the condensed-matter literature \cite{Pozsgay}.

This interesting relationship can be extended to general excitation numbers. Let $M=L/2-2m$, then
\begin{align}
2^L \Big(\frac{i}{2}\Big)^M(2m)!|\MPS \rangle \Big|_{M=L/2-2m}= (S^+)^{2m}|\Neel\rangle  + S^- |\ldots\rangle\eqndot
\end{align}
The state $(S^+)^{2m}|\Neel\rangle$ is called the $(2m)$-raised N\'eel state \cite{Brockmann}. This means that the sought-after one-point functions can be rewritten in terms of a condensed-matter problem and the results from the condensed-matter literature then provide proofs of the formulas that we just presented above.\footnote{See also \cite{Foda:2015nfk} for an alternative proof.}

\subsubsection{General \texorpdfstring{$k$}{k}}

The one-point function for general $k$ can be derived from the case $k=2$ in a recursive way. This is due to the fact that there is a recursive relation between matrix product states with different values of $k$: 
\begin{align}\label{eq:recurMPS}
|\mathrm{MPS}\rangle_{k+2} = T_1({\textstyle\frac{ik}{2}}) \, |\mathrm{MPS}\rangle_k - \left(\frac{k+1}{k-1}\right)^L |\mathrm{MPS}\rangle_{k-2}\eqncom
\end{align}
where $k\geq2$ and $|\mathrm{MPS}\rangle_0 = |\mathrm{MPS}\rangle_1=0$. 

Here, $T_1(v)$ is the transfer matrix of the XXX$_{1/2}$ Heisenberg spin chain, see J.L.\ Jacobsen's lecture:\footnote{This R-matrix is related to the one in J.L.\ Jacobsen's notes by a rescaling and by taking the appropriate $\cos\gamma \rightarrow1$ limit.}
\begin{align}
T_1(v)  := \mathrm{tr}_a(R_{aL} \ldots R_{a1} )\eqncom
\end{align}
with the R-matrix
\begin{align}
R(v) = 1 + \frac{ \permop}{v-\frac{i}{2}}\eqncom
\end{align}
which is expressed in terms of the permutation operator $\permop$. As usual, the label $a$ refers to an auxiliary $2$-dimensional space, $\mathbb{C}^2$, which is traced over in the definition of 
$T_1(v)$. 

The idea behind the proof of formula~(\ref{eq:recurMPS}) is to consider the local action of the R-operator. 
The matrix product state is formed out of the local building blocks
\begin{align}
\Big( t_1^{(k)} \otimes | \!\uparrow\rangle+ t_2^{(k)}\otimes | \!\downarrow\rangle \Big) \in \mathbb{C}^2 \otimes \mathrm{GL}(\mathbb{C}^k)\eqndot
\end{align}
Now, we add an additional auxiliary $\mathbb{C}^2$ space and consider the action of $R$ on the physical space which gives
\begin{align}\nonumber
R_{ia}({\textstyle\frac{ik}{2}}) \left[ \left\langle \uparrow_i \right| \otimes t_1^{(k)} +\left\langle \downarrow_i \right|\otimes t_2^{(k)}  \right]  =: 
\left(\left\langle \uparrow_i \right|\otimes \tau_1^{(k)} +\left\langle \downarrow_i \right|\otimes \tau_2^{(k)}\right)
\in \mathbb{C}^2 \otimes \mathrm{GL}(\mathbb{C}^{2k})\eqncom
\end{align}
where the matrices $\tau_{1,2}^{(k)}$ are given by
\begin{align}
&\tau_1^{(k)} = 
\begin{pmatrix}
\frac{k+1}{k-1} t^{(k)}_1 & 0 \\
\frac{2}{k-1}t^{(k)}_2 & t^{(k)}_1
\end{pmatrix}\eqncom
&&\tau_2^{(k)} = 
\begin{pmatrix}
t_2^{(k)} & \frac{2}{k-1}t_1^{(k)} \\
0 & \frac{k+1}{k-1}t_2^{(k)}
\end{pmatrix}\eqndot
\end{align}
The important observation is now that there exists a similarity transformation $A$ such that
\begin{align}
A \tau_i^{(k)} A^{-1} = \begin{pmatrix}
t_i^{(k+2)} & 0 \\
\star_i & \frac{k+1}{k-1}t_i^{(k-2)}
\end{pmatrix}\eqncom
\end{align}
where $\star_i$ stands for some irrelevant non-trivial entries~\cite{Buhl-Mortensen:2015gfd}.
This relation immediately proves the recursion relation~(\ref{eq:recurMPS}). 
\begin{exc} 
Check that the transformation $U=\left(
\begin{smallmatrix}
 1 & 0 & \sqrt{3} & 0 \\
 0 & \sqrt{3} & 0 & 1 \\
 i & 0 & -i \sqrt{3} & 0 \\
 0 & i \sqrt{3} & 0 & -i \\
\end{smallmatrix}
\right)$ identifies $\tau_{1,2}^{(4)} \sim t^{(4)}_{1,2}$.
\end{exc}

As discussed in J.L.\ Jacobsen's lecture notes, the Bethe states  $|\textbf{u}\rangle$ are eigenvectors of the transfer matrix with eigenvalues
\begin{align}
\Lambda(v|\textbf{u})  =  \left(\frac{v+\frac{i}{2}}{v-\frac{i}{2}}\right)^L  \prod_{i=1}^M \frac{v-u_i-i}{v-u_i} +\prod^M_{i=1} \frac{v-u_i+i}{v-u_i}\eqndot
\end{align}
The recursion relation \eqref{eq:recurMPS} then fixes all overlap functions $C_k$ for even $k$ in terms of $C_2$ and $C_0\equiv0$ by the following recursion relation:
\begin{align}
C_{k+2}  = \Lambda\left(\tfrac{ik}{2}\middle|\{u_i\}\right) C_{k} - \left(\frac{k+1}{k-1}\right)^L C_{k-2}\eqndot
\end{align}
This then implies the following explicit form for the one-point function for $k>2$:
\begin{align}\label{eq:SU2genK}
C_k =i^L T_{k-1}(0)\sqrt{\frac{Q(\frac{i}{2})Q(0)}{Q^2(\frac{ik}{2})} }\sqrt{\frac{\det  G_+}{\det  G_-}} \eqncom
\end{align}
where 
\begin{align}\label{eq:Transfer}
T_n(u) =
\!\! \sum_{a=-\frac{n}{2}}^{\frac{n}{2}}\!\! (u+ia)^L\frac{Q(u+\frac{n+1}{2}i)Q(u-\frac{n+1}{2}i)}{Q(u+(a-\frac{1}{2})i)Q(u+(a+\frac{1}{2})i)}\eqndot
\end{align}
The function $T_n(u)$ can be identified as the transfer matrix of the Heisenberg spin chain where the auxiliary space is the $(n+1)$-dimensional representation.

Since the recursion relation \eqref{eq:recurMPS} goes in steps of two, the $k=2$ result extends to all even $k$. Of course, equation \eqref{eq:SU2genK} is well-defined for any $k$ and from numerical examples it is easily seen that it also works for odd $k$. Of course, by using \eqref{eq:recurMPS}, we see that for a proof of \eqref{eq:SU2genK} for odd $k$ we only need a proof for $k=3$. This is still an open question. However, there seems to be a remarkable relation between $C_2$ and $C_3$. From \eqref{eq:SU2genK}, we find
\begin{align}
C_3 = 2^L \frac{Q(0)}{Q(\frac{i}{2})}C_2\eqndot
\end{align}
This suggest that $C_3$ and $C_2$ are related by Q-operators \cite{Bazhanov:2010ts} rather than a transfer matrix, which has been checked for states with length up to 8 \cite{Buhl-Mortensen:2015gfd}.

\subsection{One-loop one-point functions in the \texorpdfstring{$SU(2)$}{SU(2)} sector\label{one-loop-one-point}}

In order to calculate quantum corrections in the defect CFT, the action \eqref{eq: SYM-action} has to be expanded around the classical solution \eqref{eq:vev}:
\begin{equation}
 \scal_i=\scalc_i+\scalq_i\eqndot
\end{equation}
As the vacuum expectation values differ among the different flavours and are given by non-diagonal matrices in colour space, this leads to a mass matrix that mixes the different flavour and colour components of the fields. This mixing problem was solved in \cite{Buhl-Mortensen:2016pxs,Buhl-Mortensen:2016jqo}. Moreover, the vacuum expectation values are proportional to the inverse distance to the defect, $1/x_3$, such that the mass eigenvalues depend on $1/x_3$ as well. Via a Weyl transformation, this $x_3$-dependence can be absorbed to obtain standard propagators in an effective (auxiliary) $AdS_4$ space \cite{Nagasaki:2011ue,Buhl-Mortensen:2016pxs,Buhl-Mortensen:2016jqo}.

\begin{figure}[t]
\centering
 \begin{subfigure}{0.3\textwidth}
\centering
  \includegraphics{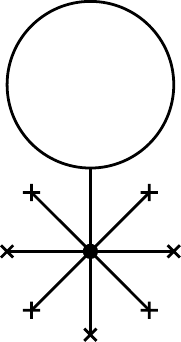}%
\caption{\phantom{.}}
  \label{subfig: lolipop}
\end{subfigure} \qquad 
 \begin{subfigure}{0.3\textwidth}
\centering
  \includegraphics{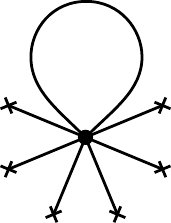}%
\caption{\phantom{.}}
  \label{subfig: tadpole}
\end{subfigure}
\caption{Two diagrams have to be considered for the one-loop correction to a one-point function: the lollipop diagram (\subref{subfig: lolipop}) and the tadpole diagram  (\subref{subfig: tadpole}).
 \label{fig: one-loop one-point functions}}
\end{figure}

At one-loop order, two different diagrams have to be considered for the one-loop correction to a one-point function of a single-trace operator built from scalars, see figure \ref{fig: one-loop one-point functions}.
The first of these diagrams, called the lollipop diagram, arises when expanding the composite operator to linear order in the quantum fields. It is given by 
\begin{equation}
\label{eq: general lollipop diagram}
 \langle\mathcal{O}\rangle_{\text{1-loop,lol}}(x)=\Psi^{i_1i_2\dots i_L}
 \sum_{j=1}^L\tr(\scalc_{i_1}\dots \contraction{}{\scalq}{_{i_j}\dots \scalc_{i_L})(x)\int\de^4y\sum_{\Phi_1,\Phi_2,\Phi_3}V_3(}{\Phi}\scalq_{i_j}\dots \scalc_{i_L})(x)\int\de^4y\sum_{\Phi_1,\Phi_2,\Phi_3}V_3(\Phi_1,\contraction{}{\Phi}{_2,}{\Phi}\Phi_2,\Phi_3)(y)\eqndot
\end{equation}
The sum in this expression is over all cubic vertices of the defect CFT, i.e.\ the original cubic vertices of $\mathcal{N}=4$ SYM theory and the additional cubic vertices that arise from inserting one scalar vacuum expectation value into the quartic vertices of $\mathcal{N}=4$ SYM theory.
As can be seen from figure \ref{subfig: lolipop}, the lollipop diagram is one-particle reducible and stems from the one-loop correction to the classical solution \eqref{eq:vev}:
\begin{align}
\label{eq: general lollipop diagram reformulated}
 \langle\mathcal{O}\rangle_{\text{1-loop,lol}}(x)&=\Psi^{i_1i_2\dots i_L}
 \sum_{j=1}^L\tr(\scalc_{i_1}\dots\langle \scal_{i_j}\rangle_{\text{1-loop}}\dots \scalc_{i_L})(x) \eqncom
 \intertext{where}
 \label{eq: general lollipop diagram reformulated 2}
 \langle \scal_i\rangle_{\text{1-loop}}(x)&=\contraction{}{\scalq}{_{i}(x)\int\de^4y\sum_{\Phi_1,\Phi_2,\Phi_3}V_3(}{\Phi}\scalq_{i}(x)\int\de^4y\sum_{\Phi_1,\Phi_2,\Phi_3}V_3(\Phi_1,\contraction{}{\Phi}{_2,}{\Phi}\Phi_2,\Phi_3)(y)
 \eqndot
\end{align}
In \cite{Buhl-Mortensen:2016pxs,Buhl-Mortensen:2016jqo}, this correction was calculated and shown to vanish provided that the employed renormalisation scheme preserves supersymmetry:
\begin{equation}
\label{eq: correction to scalar vev main text}
 \langle \scal_i\rangle_{\text{1-loop}}(x)= 0
 \eqndot
\end{equation}
This leaves us with the contribution of the diagram in figure \ref{subfig: tadpole}, called tadpole diagram.
The tadpole diagram arises from expanding the composite operator to quadratic order in the quantum fields and contracting these two quantum fields with a propagator:
\begin{equation}
\label{eq: tadpole}
 \langle\mathcal{O}\rangle_{\text{1-loop,tad}}(x)=\sum_{j_1,j_2}\Psi^{i_1\dots i_{j_1} \dots i_{j_2} \dots i_L}\tr(\scalc_{i_1}\dots \contraction{}{\scalq}{_{i_{j_1}}\dots }{\scalq}\scalq_{i_{j_1}}\dots \scalq_{i_{j_2}}\dots\scalc_{i_L})(x)\eqndot
\end{equation}

In addition to the above Feynman diagrams, the one-loop one-point function receives a contribution from the one-loop correction to the Bethe eigenstate, i.e.\ the two-loop eigenstate.
As the $SO(6)$ sector is not closed under renormalisation beyond one-loop order, we thus have to restrict ourselves to the $SU(2)$ sector, which is closed at all loop orders.
We decompose the complex scalars in the $SU(2)$ sector as 
\begin{equation}
\label{eq: decomposition into blocks}
\begin{aligned}
 X&=[X]_{n,n'}E^{n}{}_{n'}+[X]_{n,a}E^{n}{}_{a}+[X]_{a,n}E^{a}{}_{n}+[X]_{a,a'}E^{a}{}_{a'}
 \eqncom
\end{aligned}
 \end{equation}
and similarly for $Y$. The indices $n,n^\prime$ take values $1,\ldots, k$ and the indices $a,a^\prime$ run from $k+1$ to $N$. In other words, $[X]_{n,n'}$ simply corresponds to the $k\times k$ block of the $N\times N$ $U(N)$ matrix.

The number of components in the $k\times k$ block does not scale with $N$ and the components in the $(N-k)\times(N-k)$ block drop out when multiplied from the left or the right with a classical field, which is non-vanishing only in the $k\times k$ block. 
Hence, the only contribution in the large-$N$ limit stems from the components in the $k\times(N-k)$ and $(N-k)\times k$ blocks. As these drop out unless they are neighbouring, we are back at the statement that only interactions among neighbouring fields contribute in the planar limit.
Using dimensional regularisation in the $3-2\varepsilon$ directions parallel to the defect, the required propagators read \cite{Buhl-Mortensen:2016jqo}
\begin{equation}
\label{eq: equal propagator}
\begin{aligned}
 &\langle [\Xq]_{n,a}(x)[\Xq]_{a',n'}(x)\rangle=\langle [\Yq]_{n,a}(x)[\Yq]_{a',n'}(x)\rangle=\delta_{a,a'}\delta_{n,n'}\frac{g^2}{N}\frac{1}{(x_3)^2}
 \end{aligned}
\end{equation}
and
\begin{equation}
\label{eq: unequal propagator}
\begin{aligned}
 &\langle [X]_{n,a}(x)[Y]_{a',n'}(x)\rangle=-\langle [Y]_{n,a}(x)[X]_{a',n'}(x)\rangle\\
 &=\delta_{a,a'}[\comm{\Xc}{\Yc}]_{n,n'}(x)
 \frac{2g^2}{N}
 \left(
 -\frac{1}{2 \peps} - \frac{1}{2}\log(4\pi)+\frac{1}{2}\gammaE - \log (x_3)+\digamma(\tfrac{k+1}{2})
 \right)\eqndot
 \end{aligned}
\end{equation}
In these expressions, $\digamma$ denotes Euler's digamma function.
Inserting these propagators into \eqref{eq: tadpole} yields
\begin{align}
  \langle\mathcal{O}\rangle_{\text{1-loop,tad}}(x)&=g^2\frac{1}{(x_3)^2}\sum_{j}\delta_{s_j=s_{j+1}}\Psi^{s_1\dots s_j\, s_{j+1} \dots s_L}\tr(\scalc_{s_1}\dots\scalc_{s_{j-1}} \scalc_{s_{j+2}}\dots\scalc_{s_L})(x)\nonumber\\
&\phaneq +2g^2
\left(
 -\frac{1}{2 \peps} - \frac{1}{2}\log(4\pi)+\frac{1}{2}\gammaE - \log (x_3)+\digamma(\tfrac{k+1}{2}) \right)\\
 &\phaneq\phaneq\times
\sum_{j}\Psi^{s_1\dots s_j\, s_{j+1} \dots s_L}\tr(\scalc_{s_1}\dots\scalc_{s_{j-1}} [\scalc_{s_{j}},\scalc_{s_{j+1}}]\scalc_{s_{j+2}}\dots\scalc_{s_L})(x)\eqndot\nonumber
 \end{align}
We observe that the first term, stemming form \eqref{eq: equal propagator}, is finite.
The second term, stemming from \eqref{eq: unequal propagator}, however, is (ultraviolet) divergent. The divergence has to be cancelled by the renormalisation constant $\mathcal{Z}_\cO$, providing us with a second way to derive the one-loop dilatation operator \eqref{eq:HfromD}!\footnote{The ultraviolet divergence occurring at an operator depends only on the operator but not on the quantity it occurs in. This way, the renormalisation constant and thus the dilatation operator can be determined from one-point functions, two-point functions, three- and higher-point functions, from form factors etc., or simply from calculating its vertex renormalisation as one would for computing a beta function.}
Using the renormalisation constant \eqref{eq:Z-our-scheme} in the renormalisation scheme that leaves the one-loop two-point function normalised, we find
\begin{align}
  &\langle\mathcal{Z}\mathcal{O}\rangle_{\text{1-loop,tad}}(x)=g^2\frac{1}{(x_3)^2}\sum_{j}\delta_{s_j=s_{j+1}}\Psi^{s_1\dots s_j\, s_{j+1} \dots i_L}\tr(\scalc_{s_1}\dots\scalc_{s_{j-1}} \scalc_{s_{j+2}}\dots\scalc_{s_L})(x)\nonumber\\
 &\quad+g^2
 \left(
 \frac{1}{2} - \log2+\gammaE - \log (x_3)+\digamma(\tfrac{k+1}{2})
 \right)
\Delta^{(1)}\langle\mathcal{O}\rangle_{\text{tree}}(x)
\label{eq: field theory result for the one-point function}
 \end{align}
for a one-loop eigenstate with one-loop anomalous dimension $\Delta^{(1)}$.
The term proportional to $\log (x_3)$ accounts for the correction to the scaling dimension expected from \eqref{eq: form of the one-point function}, whereas the other terms contribute to the correction to the coefficient $a$ respectively $C$.
While the second term in \eqref{eq: field theory result for the one-point function} is simply proportional to the tree-level one-point function, i.e.\ the overlap between the Bethe eigenstate and the MPS, the first term in \eqref{eq: field theory result for the one-point function} can be written as the overlap of the Bethe eigenstate with an amputated matrix product state (AMPS).

On the integrability side, the one-loop one-point function thus requires to calculate the overlap of the Bethe eigenstate with the AMPS and the overlap of the loop-corrected Bethe state \cite{Gromov:2012uv} with the MPS. This calculation was done in \cite{Buhl-Mortensen:2017ind} and shown to agree with the following conjecture for an all-loop asymptotic one-point function formula proposed there as well:
\begin{align}\label{eq:Ansatz}
C_k =  i^L\tilde{T}_{k-1}(0)\sqrt{\frac{Q(\ihalf)Q(0)}{Q^2(\frac{ik}{2})} }\sqrt{\frac{\det  \tilde{G}_+}{\det  \tilde{G}_-}} 
\,\mathbb{F}_k\eqncom
\end{align}
where the Bethe roots are assumed to satisfy the all-loop asymptotic Bethe equations \eqref{AsBetheeqns}, which are also used to define $\tilde{G}_\pm$ in analogy to \eqref{eq: definition of G pm}, and 
\begin{equation}\label{eq:Transfer quantum}
\tilde{T}_n(u) = g^L \sum_{a=-\frac{n}{2}}^{\frac{n}{2}} x(u+ia)^L
\frac{Q(u+\frac{n+1}{2}i)Q(u-\frac{n+1}{2}i)}{Q(u+(a-\frac{1}{2})i)Q(u+(a+\frac{1}{2})i)}\eqncom
\end{equation}
is the quantum transfer matrix. Moreover, the introduction of a flux factor $\mathbb{F}_k$ was needed in \eqref{eq:Ansatz}, and it was found to be of the form
\begin{align}
\mathbb{F}_k = 1+g^2 \Big[ \Psi(\textstyle{\frac{k+1}{2}}) + \gammaE - \log 2 \Big] \Delta^{(1)}+O(g^4)\,.
\end{align}
The generalisation of $\mathbb{F}_k$ to higher loops constitutes an open problem.

\section{Outlook } 
\label{Discussion} 

 Integrability continues to reveal itself in connection with yet more observables of $\mathcal{N}=4$ SYM theory.  At the time of the
major review~\cite{Beisert:2010jr},  the integrability of the spectral problem was  well understood and traces of
integrability had been spotted in the form of a Yangian symmetry for tree-level and one-loop scattering 
amplitudes~\cite{Drummond:2009fd,Beisert:2010gn}.  
Since then, the integrability properties of scattering amplitudes have been further elaborated via the introduction of a spectral parameter~\cite{Ferro:2012xw,Ferro:2013dga,Chicherin:2013ora,Broedel:2014pia}. Furthermore, integrability techniques
have been applied to the study of polygonal Wilson loops~\cite{Basso:2013vsa,Basso:2013aha}, dual to planar scattering amplitudes, as well as to smooth 
Maldacena-Wilson loops~\cite{Muller:2013rta}. Form factors of $\mathcal{N}=4$ SYM theory have been studied within
the integrability language as well, both at weak coupling~\cite{Frassek:2015rka} and at strong
coupling~\cite{Maldacena:2010kp,Gao:2013dza}. In addition,  it has been demonstrated that 
the Hagedorn temperature of $\mathcal{N}=4$ SYM~\cite{Harmark:2017yrv} 
can be obtained within 
the integrability framework.
Finally, the tools of integrability inherited from the planar spectral problem
have been exploited in the calculation of higher-point correlation functions, more precisely of three-point 
functions~\cite{Escobedo:2010xs,Escobedo:2011xw,Gromov:2012uv} and of four-point functions~\cite{Eden:2016xvg,Basso:2017khq,Bargheer:2017eoz}. These efforts have
recently been boosted by the development of the so-called hexagon techniques~\cite{Basso:2015zoa,Basso:2015eqa,Basso:2017muf} 
covered in the lectures by 
S.\ Komatsu.

  In these lectures, we have chosen to focus
on the calculation of one-point functions in a certain defect version of  $\mathcal{N}=4$ SYM theory, which  constitutes yet another novel arena for the application of  integrability methods.  Combining the tools of integrability with field-theoretical computations,
we have arrived at a closed expression for the tree-level and one-loop one-point functions of the defect version of 
$\mathcal{N}=4$ SYM theory dual to the D5-D3 probe-brane system with flux. 

 It would be interesting to carry out an integrability analysis of the one-point functions of the probe-brane system as well. For instance, one could imagine that an analysis of the classical equations of motion of open strings attached at one end to the boundary of AdS and at the other end to the probe brane would reveal some known or unknown system of integrable differential equations. 
So far, only a single classical solution of this type has been found, namely a solution corresponding to a point-like
string~\cite{Nagasaki:2012re,Buhl-Mortensen:2015gfd}. It should be possible to find solutions corresponding to various types
of spinning strings as well (for a review see e.g.~\cite{Tseytlin:2010jv}), solutions which would correspond to non-protected operators in the field-theory language.

Obviously, a pressing question on the field-theory side is whether the
closed one-point function formulas obtained at the two leading orders in perturbation theory can be extended to higher
loop orders as in the case of the spectral problem. In~\cite{Buhl-Mortensen:2017ind}, we took the first step in generalising
the results to higher loop orders by proposing an asymptotic  all-loop formula (\ref{eq:Ansatz}), where in analogy with the situation for the spectral problem  the Zhukovski transformation~(\ref{xu}) plays a key role. 
This
formula agrees with a string-theory prediction~\cite{Nagasaki:2012re} in a double-scaling parameter up to wrapping order but much more work is
needed to confirm or possibly adjust the formula.

Another obvious question is whether integrability extends to other observables in the defect set-up. So far only a few
examples of other observables have been studied, namely some
Maldacena-Wilson loops~\cite{Nagasaki:2011ue,deLeeuw:2016vgp,Aguilera-Damia:2016bqv} and two-point functions 
involving either only BPS operators~\cite{deLeeuw:2017dkd}  or one operator of length two~\cite{Widen:2017uwh}. 
As mentioned
earlier, the one- and two-point functions of the defect theory might have the virtue of providing input to the boundary conformal
bootstrap programme~\cite{Liendo:2012hy,Gliozzi:2015qsa,Billo:2016cpy}.

There exists another defect version of $\mathcal{N}=4$ SYM theory which like the present one is dual to a probe-brane system with
flux, more precisely a non-supersymmetric D7-D3 probe-brane set-up~\cite{Myers:2008me}. For this set-up, only tree-level one-point functions have been
considered and so far signs of integrability have not been observed~\cite{deLeeuw:2016ofj}. It would be interesting to 
understand the apparent difference in the integrability properties of the two defect versions of 
$\mathcal{N}=4$ SYM theory  at a more fundamental  level. 

$\mathcal{N}=4$ SYM theory has a  three-dimensional somewhat close cousin, namely  ABJM theory, which is an $\mathcal{N}=6$
supersymmetric Chern Simons matter theory. For this theory, the planar spectral problem is likewise integrable and the theory has a holographic dual in the form of type IIA string theory on 
$AdS_4\times CP^3$, see e.g.~\cite{Klose:2010ki} for a review. In analogy with the $AdS_5/CFT_4$ situation, the dual string in the ABJM case allows for certain probe-brane systems with 
fluxes~\cite{Hohenegger:2009as,Gaiotto:2009tk,Ammon:2009wc}.
It would be interesting to study to which extent defect conformal field theories result from these brane constructions and if so carry through an analysis of their one-point functions and possibly reveal novel
integrability structures.

\vspace*{1.0cm}
\paragraph{Acknowledgements}
C. Kristjansen thanks the organizers of the Les Houches Summer School
\ifarxiv
entitled
``INTEGRABILITY: FROM STATISTICAL SYSTEMS TO GAUGE THEORY''
\fi
for the
invitation to lecture there and for creating an inspiring environment.
All authors thank the organizers of the school for the
invitation to contribute to the proceedings. We furthermore thank  I.\ Buhl-Mortensen, G.\ Linardopoulos, S.\ Mori, G.\ Semenoff, K.E.\ Vardinghus, and K.\ Zarembo  for useful discussions. 
The authors were  supported   by  DFF-FNU  through grant number
 DFF-4002-00037.

\bibliographystyle{utphys2}
\bibliography{Review}

\end{fmffile}
\end{document}